\documentclass[iop]{emulateapj}
\usepackage{ifpdf}

\shorttitle{Area Expansion in Active Regions}
\shortauthors{Dud\'ik et al.}

\begin{document}

\title{On the Area Expansion of Magnetic Flux-Tubes in Solar Active Regions}

\author{Jaroslav Dud\'ik\altaffilmark{1}}
    \affil{RS Newton International Fellow, DAMTP, CMS, University of Cambridge, Wilberforce Road, Cambridge CB3 0WA, United Kingdom}
    \email{J.Dudik@damtp.cam.ac.uk}

\author{Elena Dzif\v{c}\'akov\'a}
\affil{Astronomical Institute of the Academy of Sciences of the Czech Republic, Fri\v{c}ova 298, 251 65 Ond\v{r}ejov, Czech Republic}
\email{elena@asu.cas.cz}

\and

\author{Jonathan W. Cirtain}
\affil{NASA Marshall Space Flight Center, VP 62, Huntsville, AL 35812, USA}

\altaffiltext{1}{DAPEM, Faculty of Mathematics Physics and Computer Science, Comenius University, Mlynsk\'a Dolina F2, 842 48 Bratislava, Slovakia}

\begin{abstract}
 We calculated the 3D distribution of the area expansion factors in a potential magnetic field extrapolated from the high-resolution \textit{Hinode}/SOT magnetogram of a quiescent active region NOAA 11482. Retaining only closed loops within the computational box, we show that the distribution of area expansion factors show significant structure. Loop-like structures characterized by locally lower values of the expansion factor are embedded in a smooth background. These loop-like flux-tubes have squashed cross-sections and expand with height. The distribution of the expansion factors show overall increase with height, allowing an active region core characterized by low values of the expansion factor to be distinguished. The area expansion factors obtained from extrapolation of the SOT magnetogram are compared to those obtained from an approximation of the observed magnetogram by a series of 134 submerged charges. This approximation retains the general flux distribution in the observed magnetogram, but removes the small-scale structure in both the approximated magnetogram and the 3D distribution of the area expansion factors. We argue that the structuring of the expansion factor can be a significant ingredient in producing the observed structuring of the solar corona. However, due to the potential approximation used, these results may not be applicable to loops exhibiting twist neither to active regions producing significant flares.
\end{abstract}


%
\section{Introduction}
\label{Sect:1}

Coronal loops are the basic building blocks of the solar corona. They are arch-like structures of locally denser plasma at temperatures of the order of 10$^6$\,K, emitting strongly in the X-rays, EUV and UV parts of the spectrum. The temperatures and densitities of the loops correspond to the emitting plasma being in the collisional regime and thus highly ionized. Because of the high electric conductivity, the plasma is frozen-in, and has to follow the magnetic field.

Analysis of the data obtained by coronal imaging instruments (\textit{Yohkoh}/SXT and \textit{TRACE}) showed that the cross-section of coronal loops do not appreciably expand with height \citep[e.g.,][]{Klimchuk92,Klimchuk00,Watko00,Aschwanden05,Brooks07}. This led \citet{Klimchuk00} to conclude that the loops should have circular cross-sections. The absence of loop cross-section expansion is puzzling, especially given that loops should follow the magnetic field, and that the cross-sections of magnetic flux-tubes do expand with height \citep[see e.g.,][for recent examples]{Dudik11,Peter12,Asgari12,Asgari13,Mikic13,Malanushenko13}. This is an important issue, since the geometrical effects connected with the expanding cross-section will significantly influence the interpretation of observations \citep[e.g.,][]{Vesecky79,DeForest07,Malanushenko13,Guarrasi14}, as well as the modeling of active region coronae \citep[e.g.,][]{Warren10,Dudik11}. Attempts to explain the constant loop cross-sections with coronal currents or the effects of kinetic pressure, gravitational stratificiation, or steady flows have been made. However, it was found that the linear force-free fields still expand about twice as much as the observed loops \citep{LopezFuentes06}, while coronal loops are not highly twisted \citep{Kwon08}, and the effects of pressure, steady flows, and stratification are likely to be negligible in the low-$\beta$ plasma \citep{Petrie06,Petrie08}.

\citet{DeForest07} pointed out that in the observations, unresolved expanding structures could appear to be non-expanding, and could also produce the observed ``super-hydrostatic'' pressure scale-heights, since the emitting volume increases along the loop. \citet{LopezFuentes08} studied the effects of the diffuse coronal background, in which the loops are embedded \citep{Cirtain05}. They showed that if the loop widths are near the instrument resolution limit, loops with expanding cross-sections should be distinguished from the non-expanding ones even in the presence of background contamination. The procedure was also able to distinguish unresolved loops from loops near the resolution limit. The question whether loops are resolved, or consist of individual unresolved strands, has been extensively investigated \citep[e.g.,][]{DelZanna03a,DelZanna03b,Aschwanden05,Patsourakos07,Schmelz09,Schmelz11a,Schmelz11b,Schmelz13,Brooks12,Brooks13,Peter13,DelZanna13,Winebarger14}. At least a portion of the observed loops seems to be well-resolved \citep{Aschwanden05,Aschwanden11,Brooks12,Brooks13,Peter13,Winebarger14}. How is it then possible that these loops are observed to be non-expanding?

Interestingly, some answers were provided by modeling. For example, \citet{Mok08}, \citet{Dudik11}, \citet{Peter12} and \citet{Lionello13} consistently find that the modeled coronal loops are non-expanding in spite of the fact that the underlying magnetic field is. \citet{Peter12} performed a 3D MHD model of an active region corona heated by braiding \citep[see also][]{Schrijver07} and showed that the apparently non-expanding cross-section of the modeled loops is a result of interplay between the density and temperature in the observations, which ``cut off'' visibility of parts of the expanding flux-tube. \citet{Dudik11} showed that the expansion factor is varying in space, i.e., from flux-tube to flux-tube, at scales of $\approx$2$\arcsec$ given by the resolution of the \textit{SOHO}/MDI magnetogram. This structuring was a necessary ingredient to obtain a match between the modeled active region morphology and the observed one at different temperatures. \citet{Malanushenko13} also found intrinsically expanding flux-tubes in the extrapolated potential magnetic fields, but pointed out that the cross-sections are never circular, but highly oblate. This strongly modifies visibility of the loops and can lead to selection effects in an analysis of the observations, as the loops with cross-section expanding along the line-of-sight are apparently less expanding, and have enhanced visibility. That could lead to selection effects in the analysis. Observations of the sheet-like structures in the solar chromosphere were reported by \citet{Judge11,Judge12}. However, it is unknown whether these could correspond to the flux-tubes with squashed cross-sections of \citet{Malanushenko13}.

The expansion factors found in the simulations are typically larger than unity, up to several tens \citep{Dudik11,Peter12,Asgari12,Asgari13,Mikic13,Lionello13}. These values are important, since they can change the thermodynamic character of the solutions to hydrodynamic equations. Namely, expanding loops are more likely to be unstable and undergo thermal non-equilibrium \citep{Mikic13}. This phenomenon is a possible explanation for warm ($\approx$\,1\,MK) loops, since apart from the modeled non-expansion, it could also explain evolution of lightcurves and cross-sectional temperature variations \citep{Lionello13}. Thermal non-equilibrium would also lead to coronal rain, which is observed to be ubiquitous in H$\alpha$ \citep{Antolin12}, occurring simultaneously on neighbouring strands, implying similar heating on these strands. Moreover, the H$\alpha$ blobs were observed to change shape and elongate as they moved downwards to the chromosphere, which could indicate tapering of the corresponding magnetic flux-tubes with decreasing height. Note that sporadic downflows resembling coronal rain were also reported in coronal EUV observations \citep{Schrijver01,UgarteUrra09,Kamio11}.

In this paper, we supplement the results of \citet{Malanushenko13} and revisit the question of the 3D spatial structure of the area expansion factor in potential fields, first addressed by \citet{Dudik11}. To do that, we use a \textit{Hinode}/SOT magnetogram, which has high spatial resolution of 0.3$\arcsec$, i.e., a factor of $\approx$6.6 better than the \textit{SOHO}/MDI. Observations are described in Sect. \ref{Sect:2}. Section \ref{Sect:3} summarizes the methods for obtaining potential magnetic field and the 3D distribution of the area expansion factors. Results are presented in Sect. \ref{Sect:4} and their implications discussed in Sect. \ref{Sect:5}. The main findings are summarized in Sect. \ref{Sect:6}.

%
   \begin{figure}[!t]
	\centering
	\includegraphics[width=8.8cm]{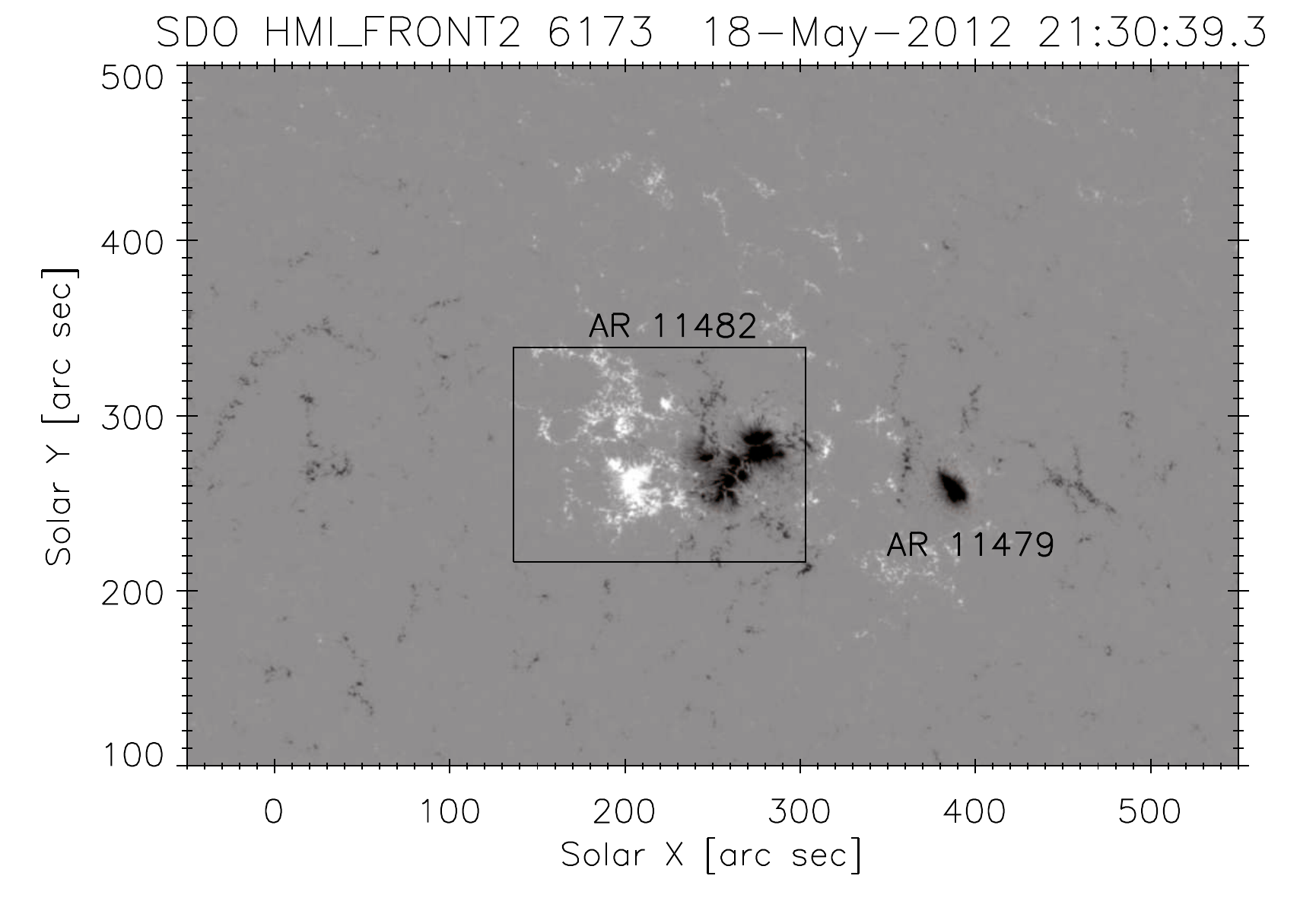}
	\includegraphics[width=8.8cm]{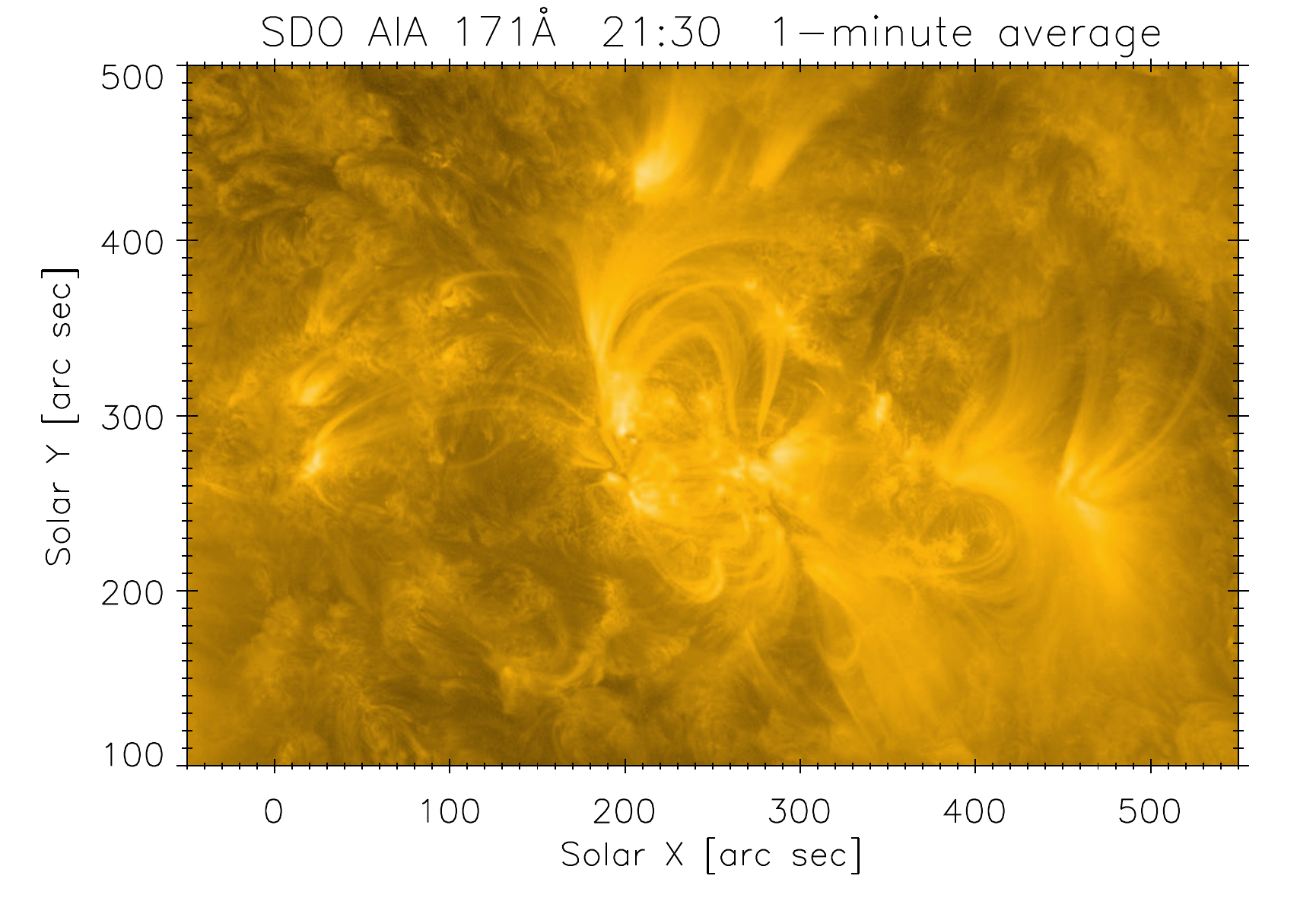}
	\includegraphics[width=8.8cm]{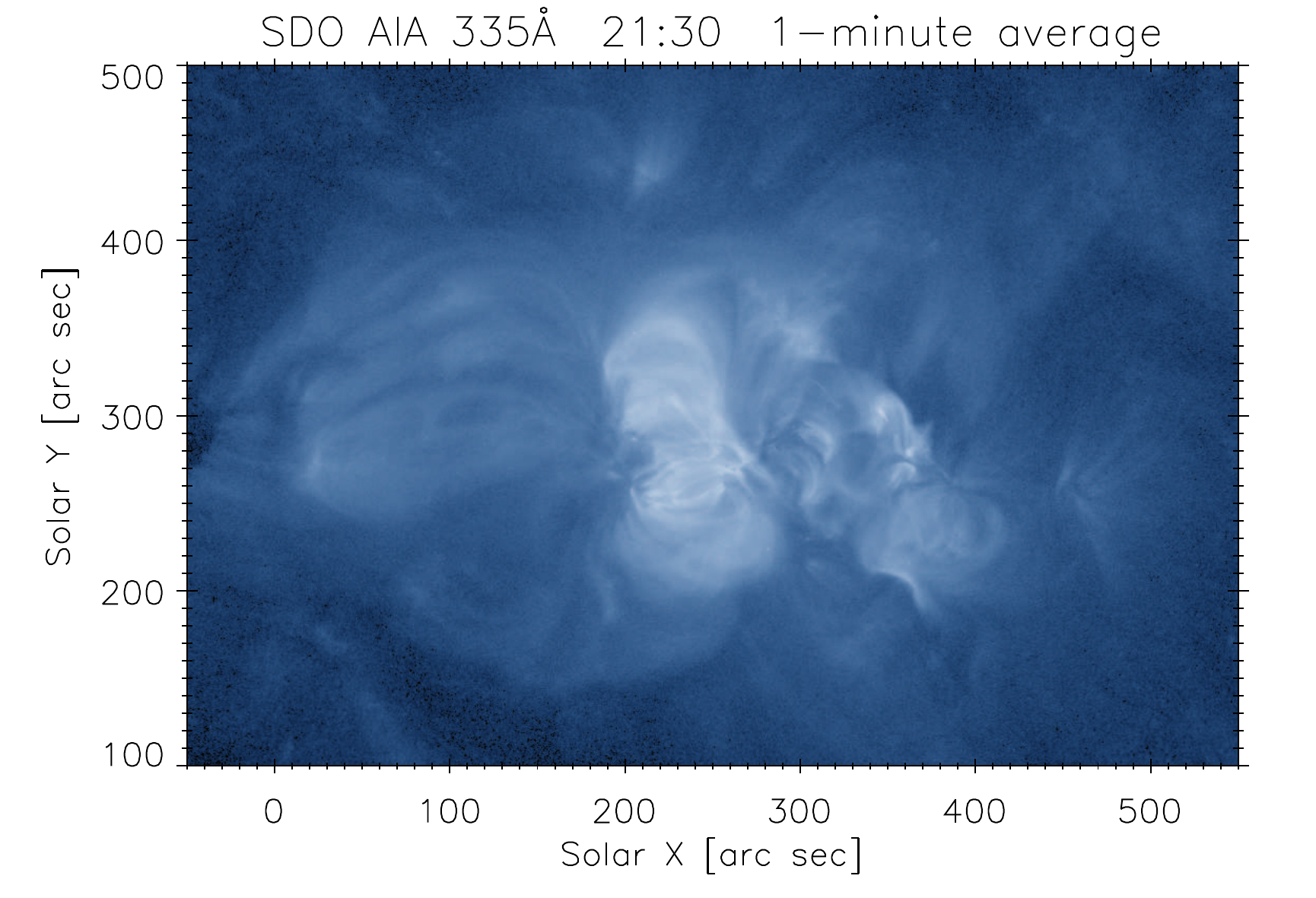}
	\caption{Context images showing the the active regions NOAA 11482 and its location with respect to the outlying quiet Sun and the AR 11479. \textit{Top}: SDO/HMI magnetogram. The locations of the two ARs are indicated. The box shows the \textit{Hinode}/SOT field-of-view. \textit{Middle} and \textit{bottom}: 1-minute average of the AIA 171\AA~and AIA 335\AA~observations, respectively. AIA counts are scaled logarithmically.}
       \label{Fig:SDO}
   \end{figure}
%
   \begin{figure*}[!ht]
	\centering
	\includegraphics[width=8.8cm]{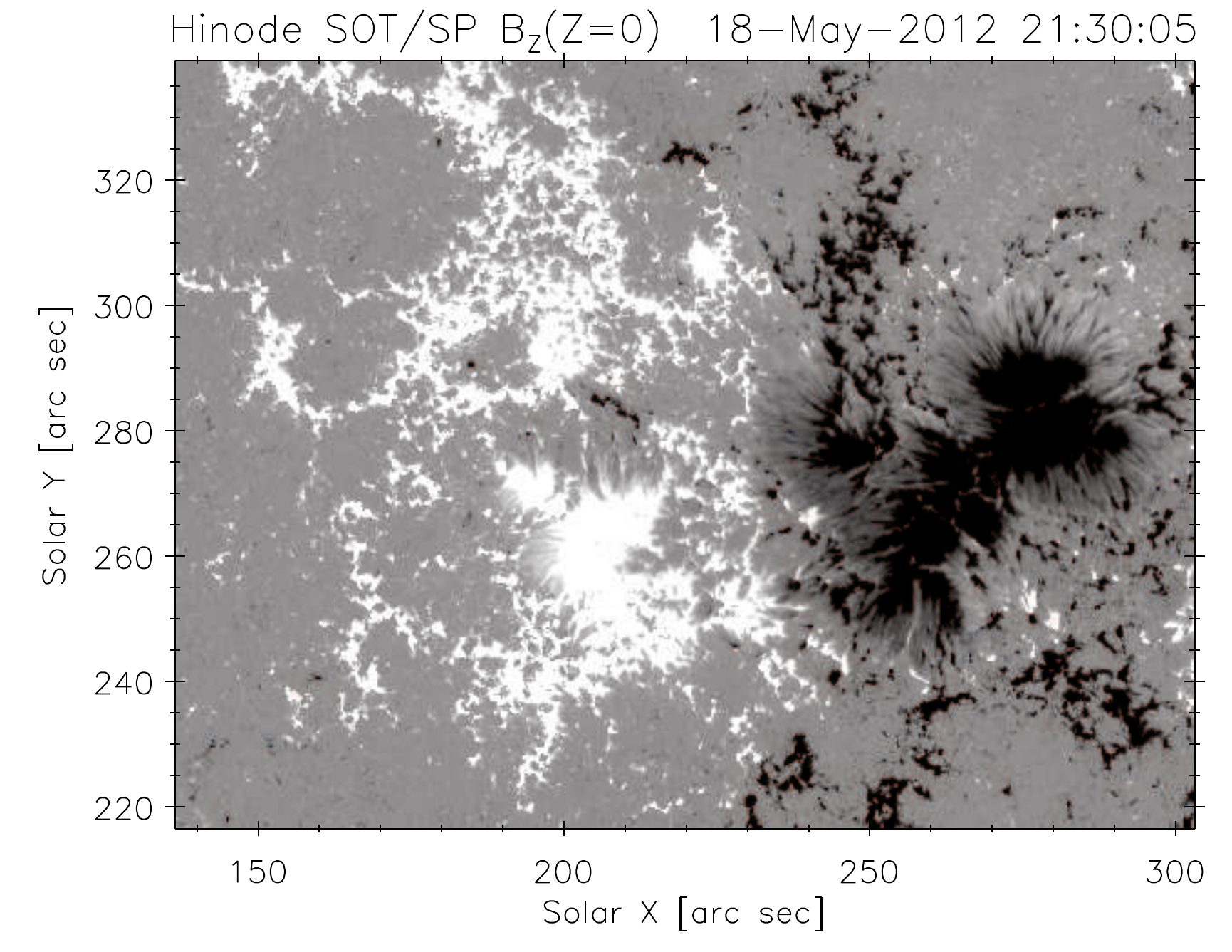}
	\includegraphics[width=8.8cm]{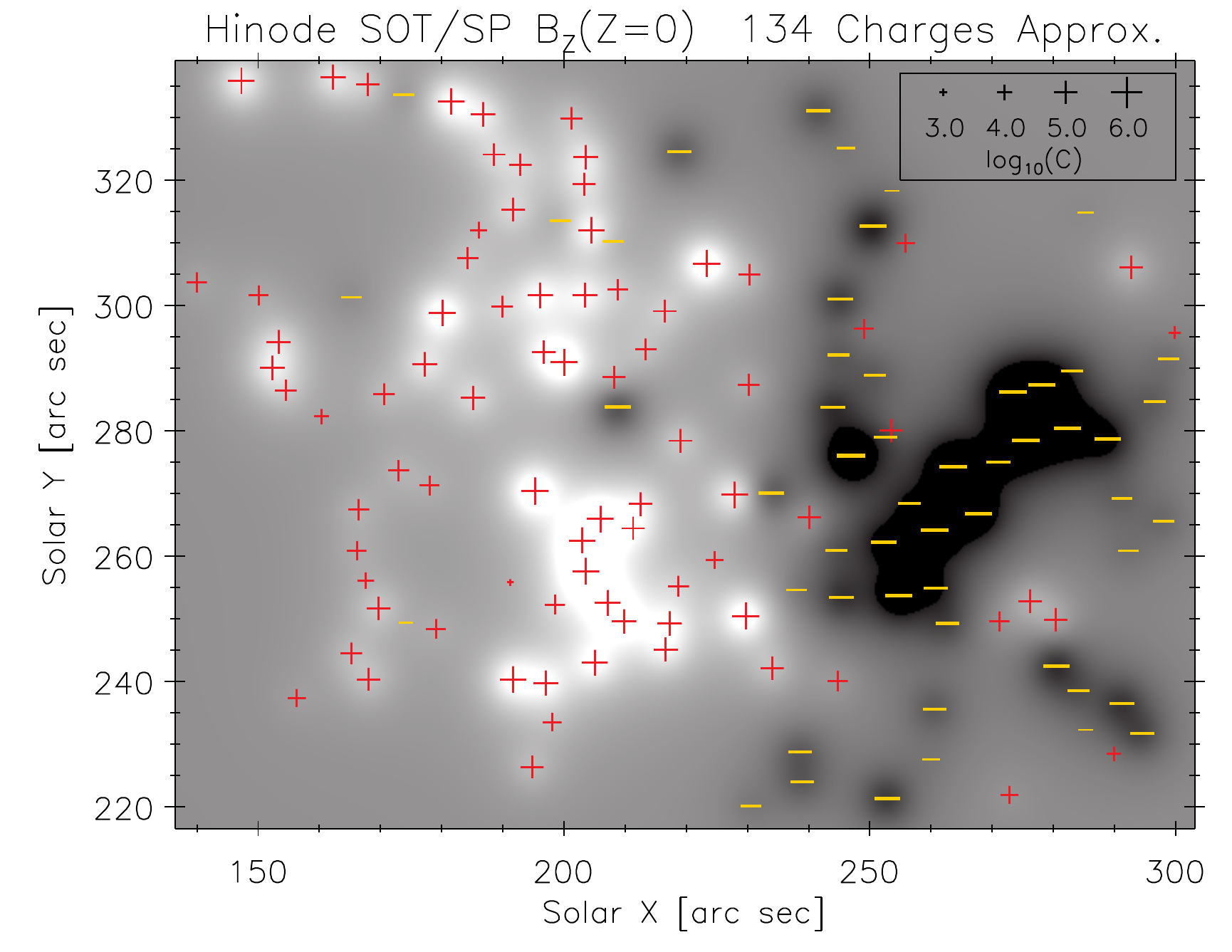}
	\includegraphics[width=8.8cm]{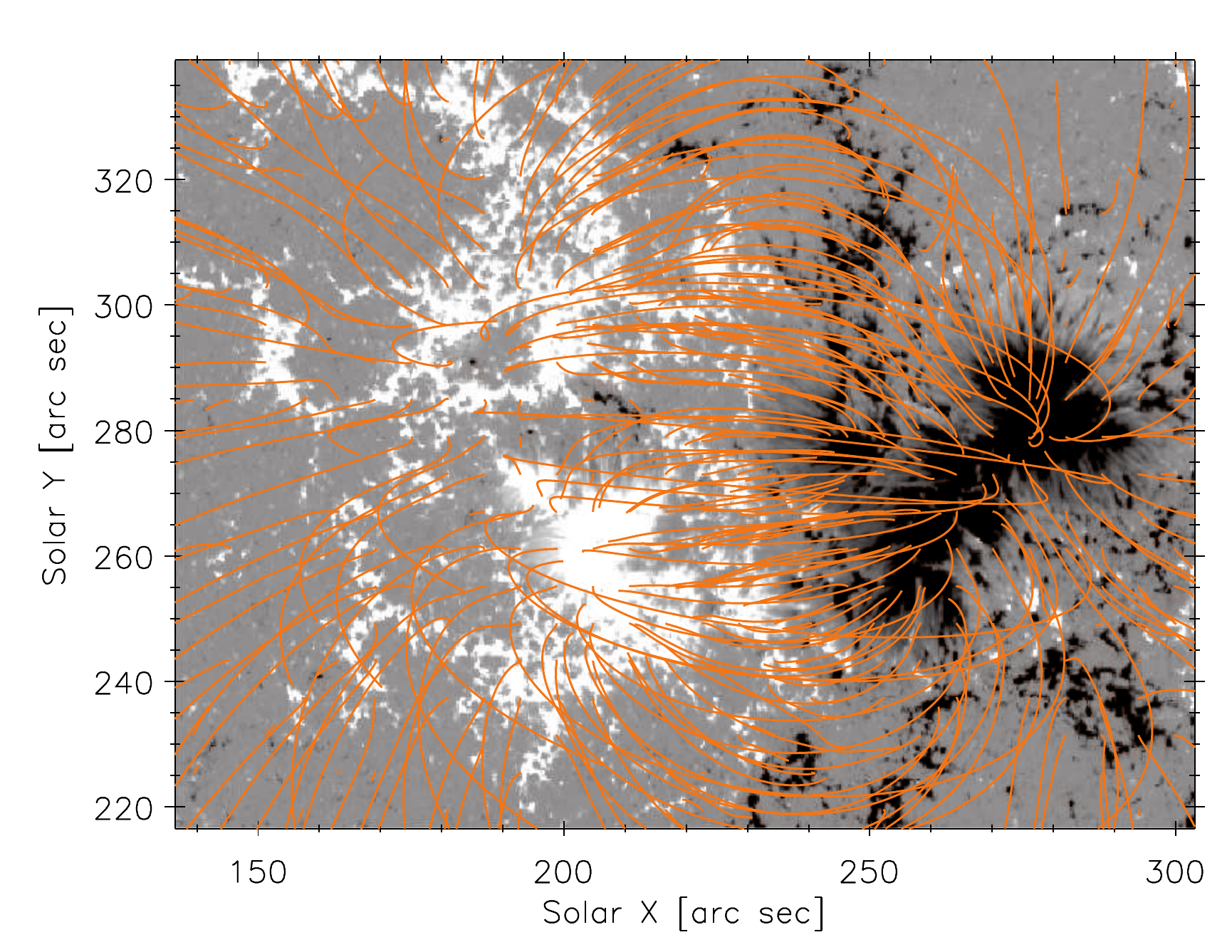}
	\includegraphics[width=8.8cm]{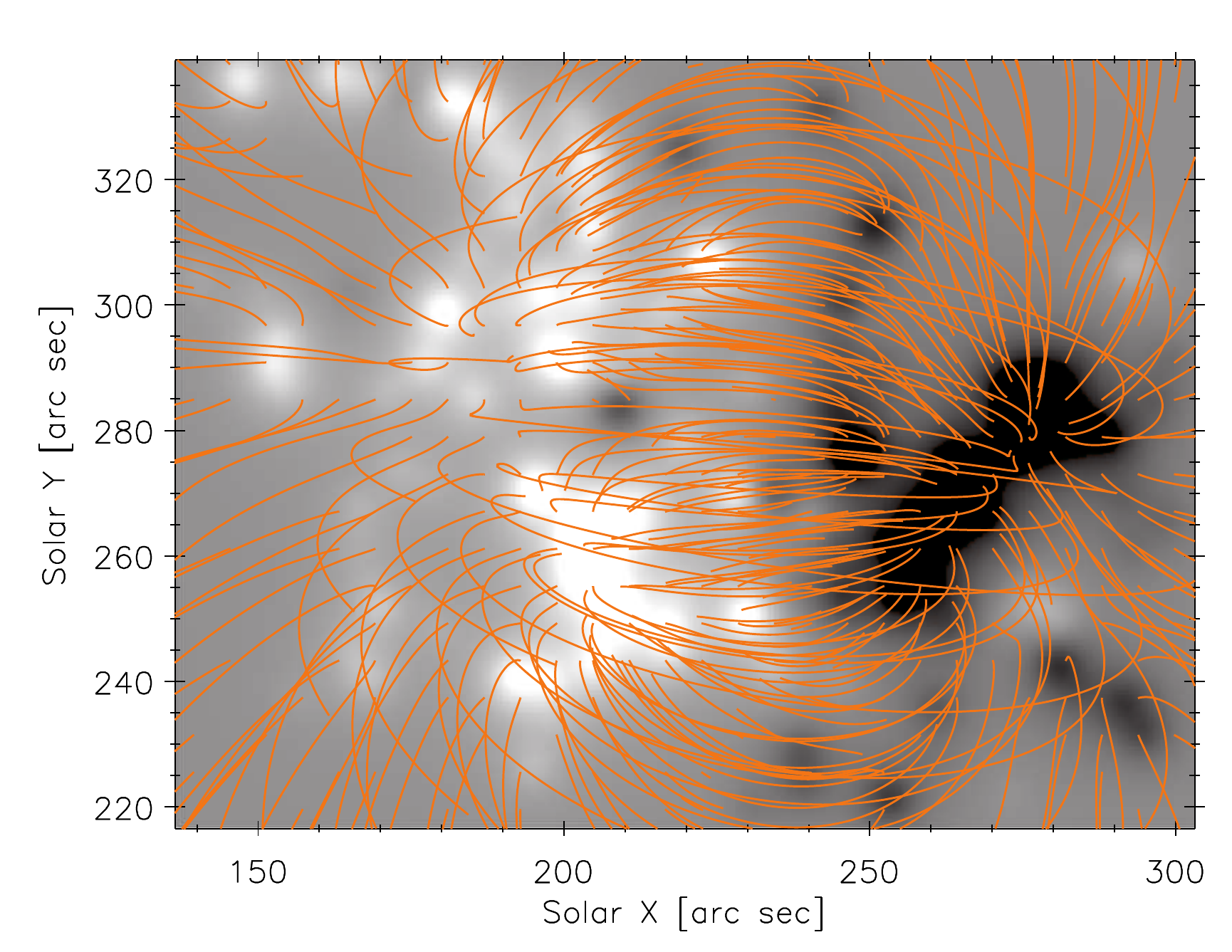}
	\includegraphics[width=8.8cm]{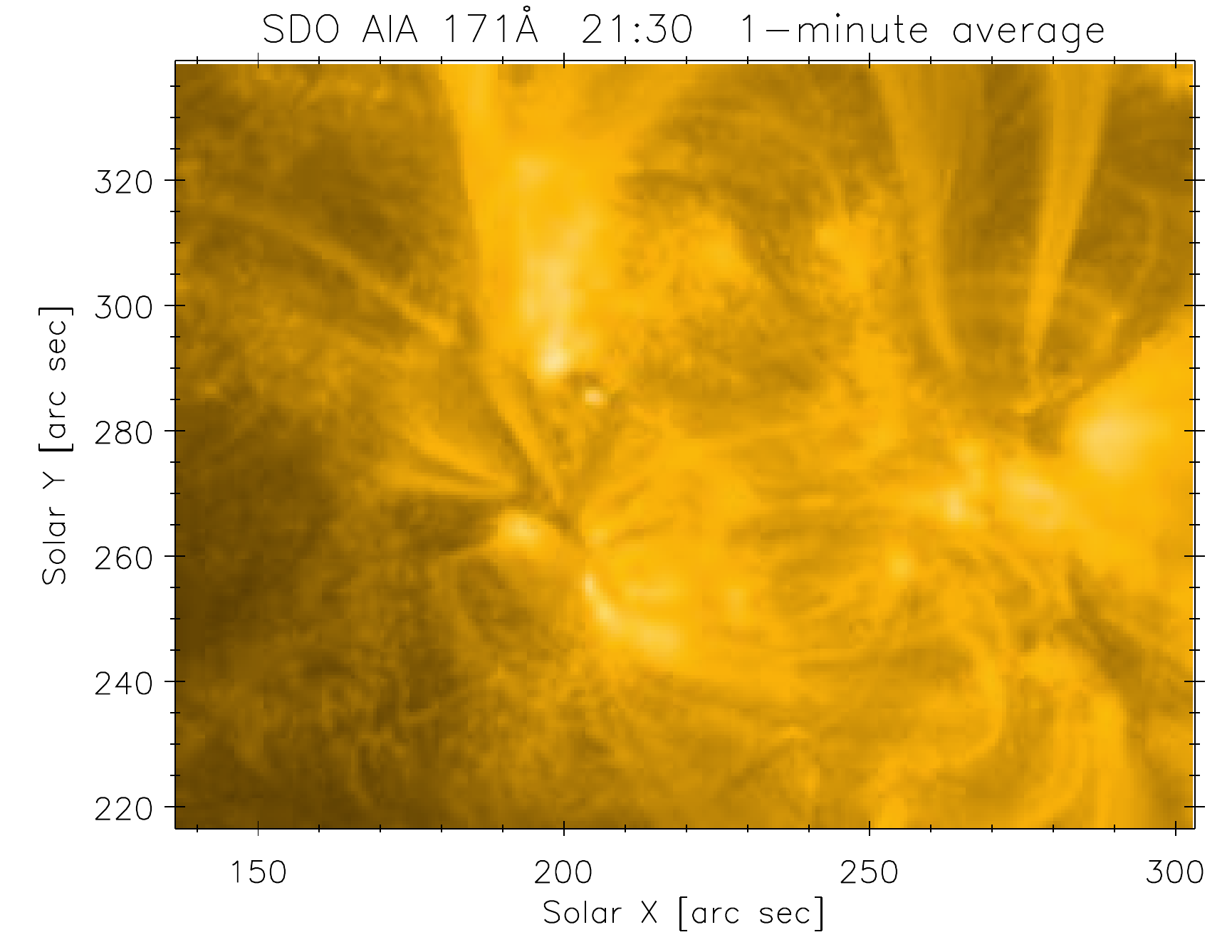}
	\includegraphics[width=8.8cm]{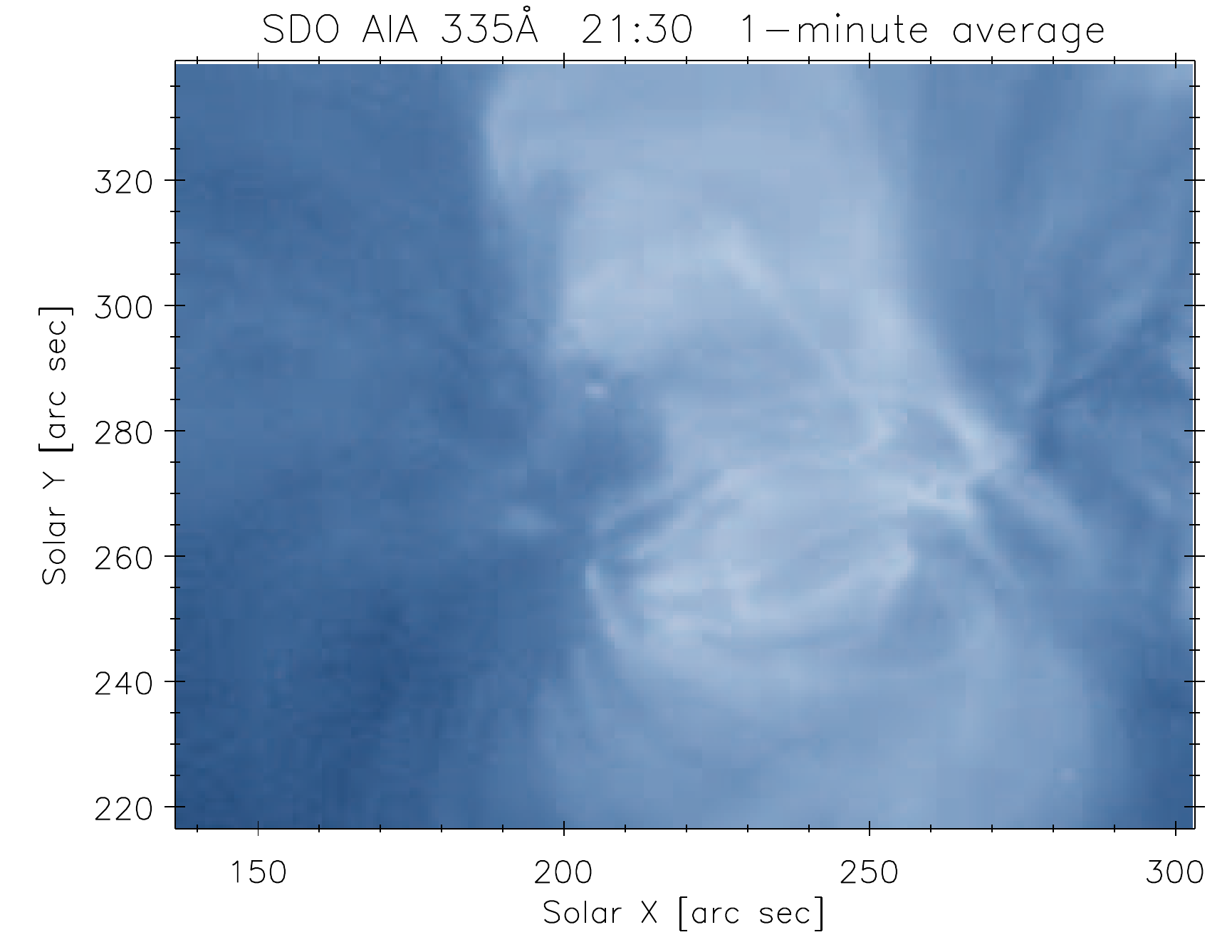}
	\caption{\textit{Top}: Hinode/SOT observations of the longitudinal magnetic field component and its approximation with 134 submerged charges. The location of the charges is denoted by the ``+'' and ``$-$'' signs. Sizes of the symbols stands for the amplitude of the charges. \textit{Middle}: Corresponding magnetic fields. \textit{Bottom}: AIA 171\AA~and 335\AA~observations of the active region corona.}
       \label{Fig:SOT}
   \end{figure*}

%
\section{Observations}
\label{Sect:2}

The Solar Optical Telescope \citep[SOT,][]{Tsuneta08,Suematsu08} is a 50\,cm diffraction-limited optical telescope onboard the Hinode satellite \citep{Kosugi07}. It is equipped with two focal-plane filter packages, a correlation tracker, as well as a spectropolarimeter (hereafter, SP). The spectropolarimeter \citep{Lites01} is an off-axis Littrow-Echelle spectrograph observing the Zeeman-sensitive \ion{Fe}{1} dual-line at 6301.5\AA~and 6302.5\AA. The observations are uninterupted, stabilized, with a spatial resolution of about 0.3$\arcsec$, spectral resolution of $\approx$21\,m\AA~and a high polarimetric accuracy of about $\approx$10$^{-3}$ \citep{Ichimoto08,Shimizu08}. The field of view is up to approximately 350$\arcsec$\,$\times$\,150$\arcsec$, which can be sufficient to encompass an active region. Both spatial resolution and the large enough field of view makes the SOT--SP instrument uniquely suited for a study of the magnetic field structure in an active region.

We performed a search of the SOT/SP level-2 database\footnote{http://sot.lmsal.com/data/sot/level2d/} containing data processed using the Milne-Eddington inversion \citep{Skumanich87,Lites90,Lites93} implemented in the MERLIN code. The criteria for a selection of a suitable active region for extrapolation (Sect. \ref{Sect:3}) were as follows:
\begin{enumerate}
 \item the active region should have simultaneous \textit{Hinode}/SOT and \textit{SDO}/AIA observations (i.e., only active regions after 2010 May 20),
 \item the active region should be bipolar, with no strong apparent departures from the potential magnetic field as observed by \textit{SDO}/AIA (i.e., no apparently twisted or S-shaped loops)
 \item the active region should contain a sunspot of each polarity, i.e., a strong flux concentration,
 \item the active region should not be located more than $\approx$30$^\circ$ away from the central meridian to minimize the projection effects, and
 \item the entire active region should be located within the SOT/SP field of view.
\end{enumerate}
Finally, we do not require that the magnetic flux of the active region is balanced (Sect. \ref{Sect:3}).

Based on the above, we selected the active region (hereafter, AR) NOAA 11482 observed by SOT on 2012 May 18, 21:30 UT. The Hale classification of this AR is $\beta \gamma / \beta \gamma$. The SOT field-of-view captures the active region almost entirely (Fig. \ref{Fig:SDO}, \textit{top}), although there are few plage polarities outside of its field-of-view. The context longitudinal magnetogram provided by the Helioseismic Magnetic Imager \citep[HMI][]{Scherrer12,Schou12a,Schou12b} onboard the Solar Dynamics Observatory \citep{Pesnell12} is shown in Fig. \ref{Fig:SDO}, \textit{top}, and the \textit{Hinode}/SOT--SP magnetogram of the AR 11482 in Fig. \ref{Fig:SOT}, \textit{top left}. The AR 11482, as observed by SOT--SP, exhibits a flux imbalance of about 0.237, calculated as the ratio of the difference between total positive an negative flux to the smaller of the two. The \textit{SDO}/HMI magnetogram indicates that the AR 11482 is located in the vicinity of the AR 11479 that exibits only one negative sunspot (Fig. \ref{Fig:SDO}, \textit{top}).

The magnetic connectivity of these active regions can be inferred using coronal loop observations made by the Atmospheric Imaging Assembly \citep[AIA,][]{Lemen12,Boerner12}. AIA observations in the 171\AA~and 335\AA~are shown in Fig. \ref{Fig:SDO}. We chose these two observations since they represent both the ``warm'' ($\approx$\,1\,MK) and ``hot'' ($>$\,2\,MK) coronal loops typically located in active regions \citep[e.g.,][]{Mason99,Klimchuk10,Reale10,Tripathi09,Tripathi11,DelZanna13}. The response of the AIA 171\AA~band peaks near log$(T/$K)\,=\,5.8, while the coronal response of the 335\AA~band peaks at log$(T/$K)\,=\,6.45 \citep[e.g.,][]{Schmelz11a,DelZanna13}. From these observations, we see that there is only one apparent, large-scale 171\AA~loop connecting connecting the trailing positive polarity of AR 11482 with the leading negative sunspot of AR 11479 (Fig. \ref{Fig:SDO}, \textit{middle}). This is not surprising, given the distance between the two polarities, and the fact that the leading negative spots of AR 11482 are located between them. The trailing positive-polarity plage of the AR 11482 is instead magnetically connected to the outlying negative quiet-Sun network polarities, located to the East of AR 11482, at approximately $X$\,=\,20$\arcsec$. This magnetic connection is observed both in AIA 171\AA~and 335\AA~(Fig. \ref{Fig:SDO}, \textit{middle} and \textit{bottom}).

We performed an approximate coalingnment between HMI and SOT that corrects for mutual offset and rotation. We determined that $\delta X$\,=\,$7.5\arcsec$, $\delta Y$\,=\,43$\arcsec$ and $\delta \phi$\,=\,+1$^\circ$. The same geometrical transformation is applied to the AIA data, which are shown in Fig. \ref{Fig:SOT}, \textit{bottom} with the field-of-view corresponding to the SOT instrument. We also note that a plate-scale correction of $\delta(\Delta X)$\,$\approx$\,1.7\% would lead to an improved match between the the location of various polarities in both datasets. However, we were not successful in obtaining close match for all polarities, mainly the small ones, even by varying $\Delta Y$. It is possible that the differences between the two datasets are a result of temporal evolution of the small polarities, or measuring different Zeeman-sensitive lines, or both. Therefore, we chose not to pursue the matter further. We consider the coalignment achieved by adjusting the mutual offset and rotation to be satisfactory for a qualitative comparison between the extrapolated magnetic field and the observed shape of the coronal loops.

We note that the AR 11482 produced only three C-class flares during its entire on-disk passage on 2012 May 12--23. Two of them, C1.3 and C1.6 occurred on May 15; the third one, C1.0, ocurred on May 19. \citet{Jeong07} studied three other active regions of comparable total magnetic flux ($\approx3.34$\,$\times$10$^22$\,Mx), ARs 10365, 10656, and 10696. These active regions showed significant total helicity and produced a number of M-class and even X-class flares during their on-disk passages. Compared to these active regions, our AR 11482 exhibits a conspicuously low number of only small flares. This could indicate a relatively low total helicity, reflected by absence of apparently twisted or S-shaped loops (Figs. \ref{Fig:SDO} and \ref{Fig:SOT}). Note also the absence of apparent twist was one of our requirements for selection of a suitable AR. Therefore, we caution the reader that the results presented in this paper (see Sects. \ref{Sect:3} and \ref{Sect:4}) may not be general and applicable to loops exhibiting twist neither to active regions producing significant flares.

\section{Magnetic field extrapolation}
\label{Sect:3}

\subsection{Choice of extrapolation method}
\label{Sect:3.1}

To study the structure of the magnetic field and its area expansion factors, we performed a potential extrapolation of the observed \textit{Hinode}/SOT magnetogram. The potential approximation ($\mathbb{\nabla} \times \mathbf{B}$\,=\,$\mathbf{0}$) is chosen because it represents the lowest approximation, and any structure therein will correspond to the state of minimum energy of the field. It is chosen also because of the lack of apparently twisted loops in the \textit{SDO}/AIA observations (Sect. \ref{Sect:2}).

Note that a mis-match between the potential extrapolation and the shape of the observed coronal loops is known to exist. This mis-match is however approximately the same for the non-linear force-free extrapolation methods \citep{Sandman09,DeRosa09}. In addition, different non-linear force-free extrapolation methods can produce different results. The reader is referred to the works of e.g., \citep{Schrijver06}, \citet{Metcalf08}, and \citet{Regnier13} for comparisons between the various non-linear force-free extrapolation methods. Both potential and non-linear force-free fields minimizing the mis-alignment with the observed loop geometry have been constructed \citep[e.g.,][]{Aschwanden10,Sandman11,Aschwanden13a,Aschwanden14,Gary14}. However, these methods use approximations of the observed magnetogram by a series of submerged magnetic monopoles (charges) or bipoles. The approximation of the flux distribution in the observed magnetogram by submerged charges will be done in Sect. \ref{Sect:3.2} and further studied in Sect. \ref{Sect:4}.

Since the AR 11482 exhibits significant flux imbalance, we performed the potential extrapolation using the Green's function method \citep{Sakurai82,Roumeliotis96}. The magnetic field in any Carthesian gridpoint above the photospheric magnetogram is given by
  \begin{equation}
	\mathbf{B}(\mathbf{r}) = \frac{1}{2\pi} \int B_Z(X,Y,Z=0) \frac{\mathbf{r} -\mathbf{R}}{\,\left|\mathbf{r} -\mathbf{R} \right|^3} \mathrm{d}X \mathrm{d}Y\,,
	\label{Eq:Green}
  \end{equation}
where $\mathbf{r} = [x,y,z]$ is the grid point location, and $\mathbf{R} = [X,Y,Z=0]$ is the Carthesian coordinate system of the observed photospheric magnetogram (Fig. \ref{Fig:SOT}, \textit{top left}). The extrapolation is performed on the entire observed SOT/SP magnetogram (562 $\times$\,413$\times$ points) up to a height of $\approx$100\,Mm, which at the resolution of 0.3$\arcsec$\,px$^{-1}$ corresponds to 432 points in the $Z$-direction.

The Green's function method (Eq. \ref{Eq:Green}) is preferred over the Fourier transform methods \citep{Nakagawa72,Alissandrakis81,Gary89}, since these assume flux-balance, and also introduce aliasing effects. The flux-balance is in fact a necessary condition in the formulation of the Fourier transform method, and is enforced by modifying the magnetogram if it is not met \citep[see also][]{Dudik08}. However, since the number of grid-points is large, $\approx$10$^8$, the Green's function method is extremely slow ($\approx$180 hours on an 8-CPU i7 machine). 

We note that the near-potentiality of the active region was one of the selection criteria (Sect. \ref{Sect:2}). From comparison of Fig. \ref{Fig:SOT}, \textit{middle left} and \textit{bottom} we see that the potential approximation represents a reasonable match to the observed shape of the coronal loops.

   \begin{figure*}[!ht]
	\centering
	\includegraphics[width=8.8cm]{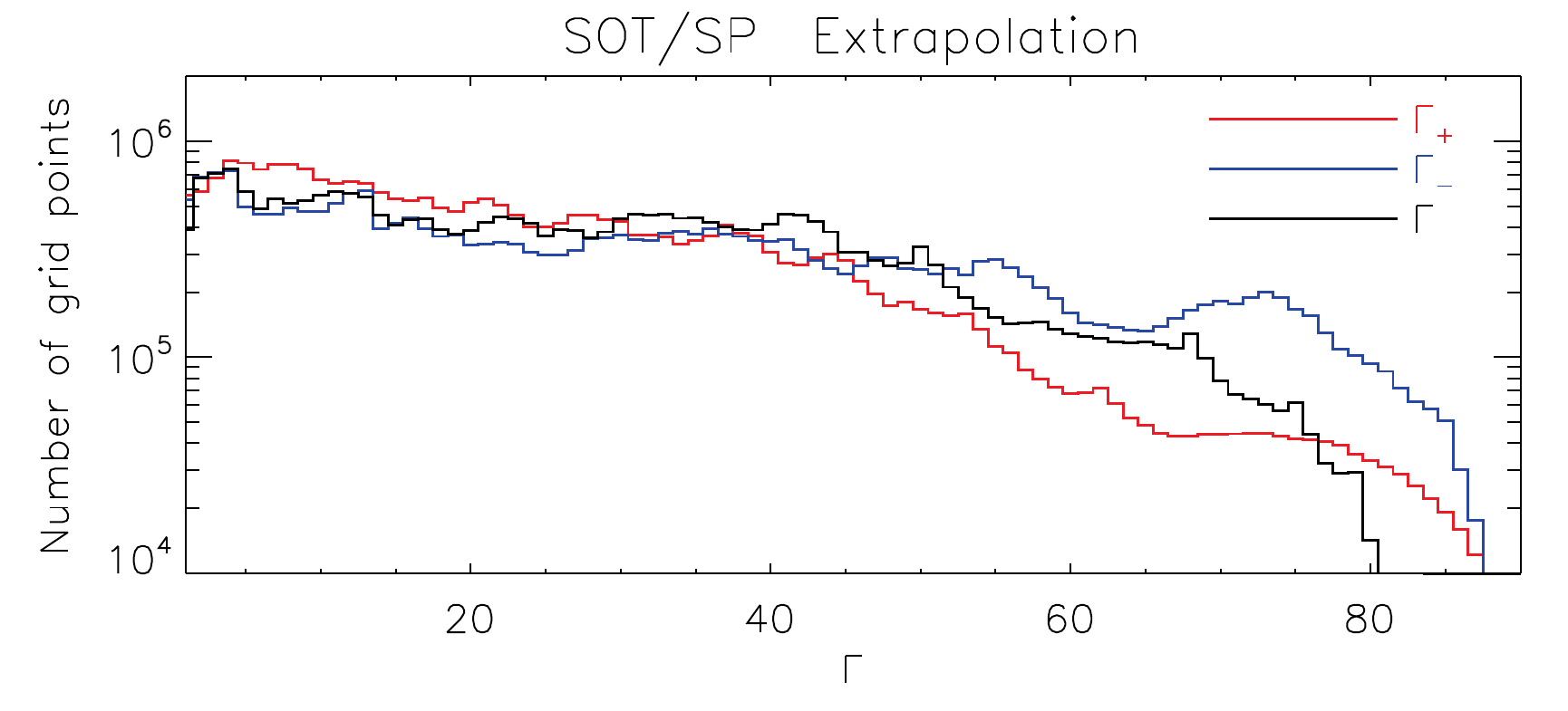}
	\includegraphics[width=8.8cm]{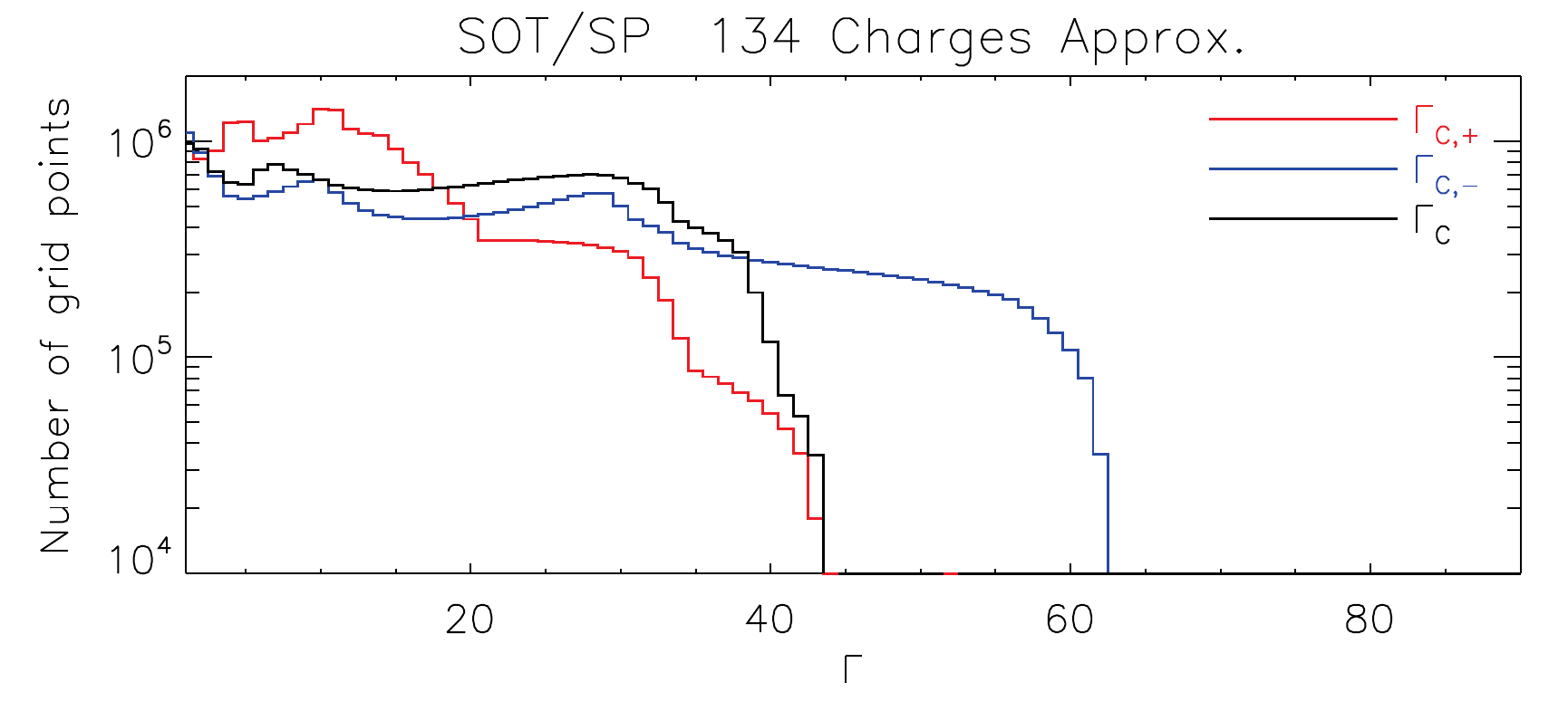}
	\includegraphics[width=8.8cm]{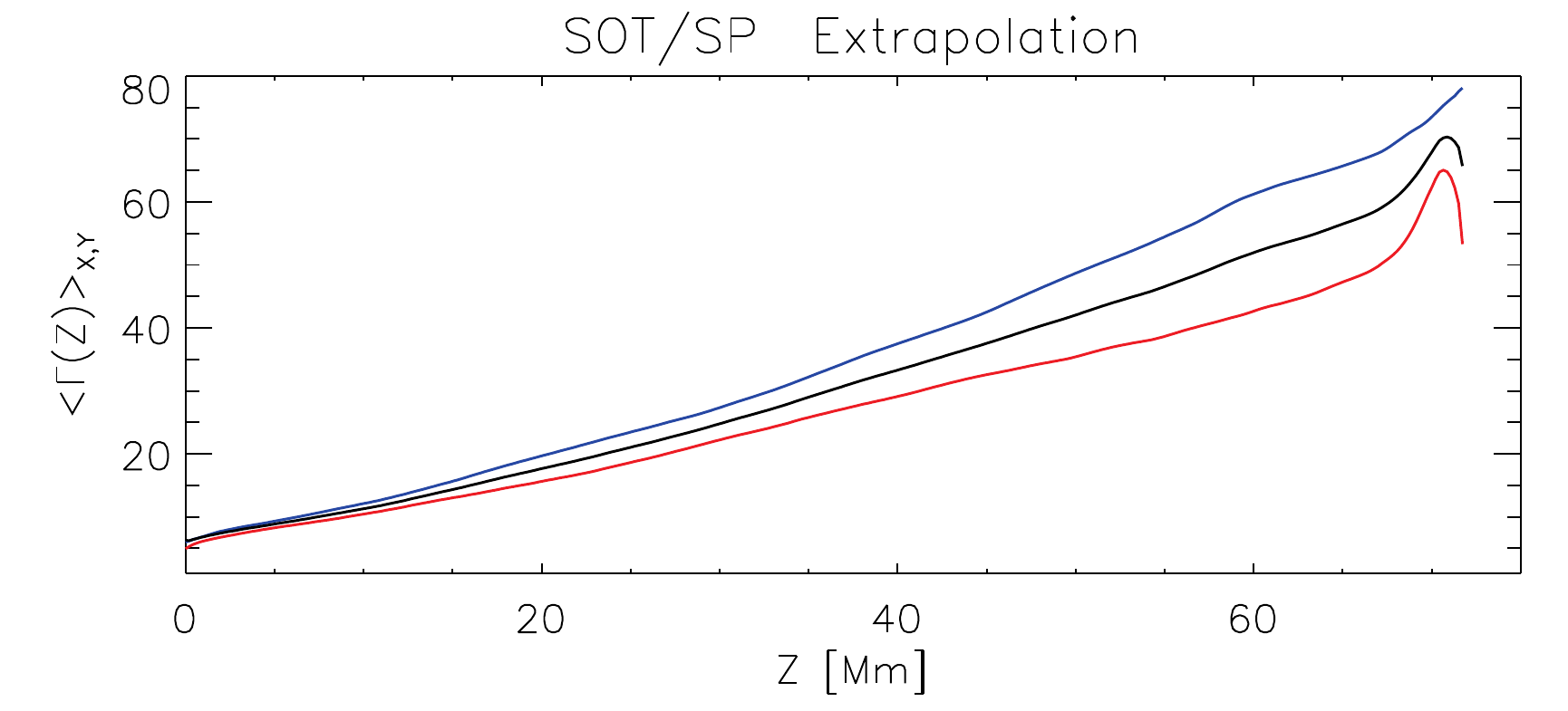}
	\includegraphics[width=8.8cm]{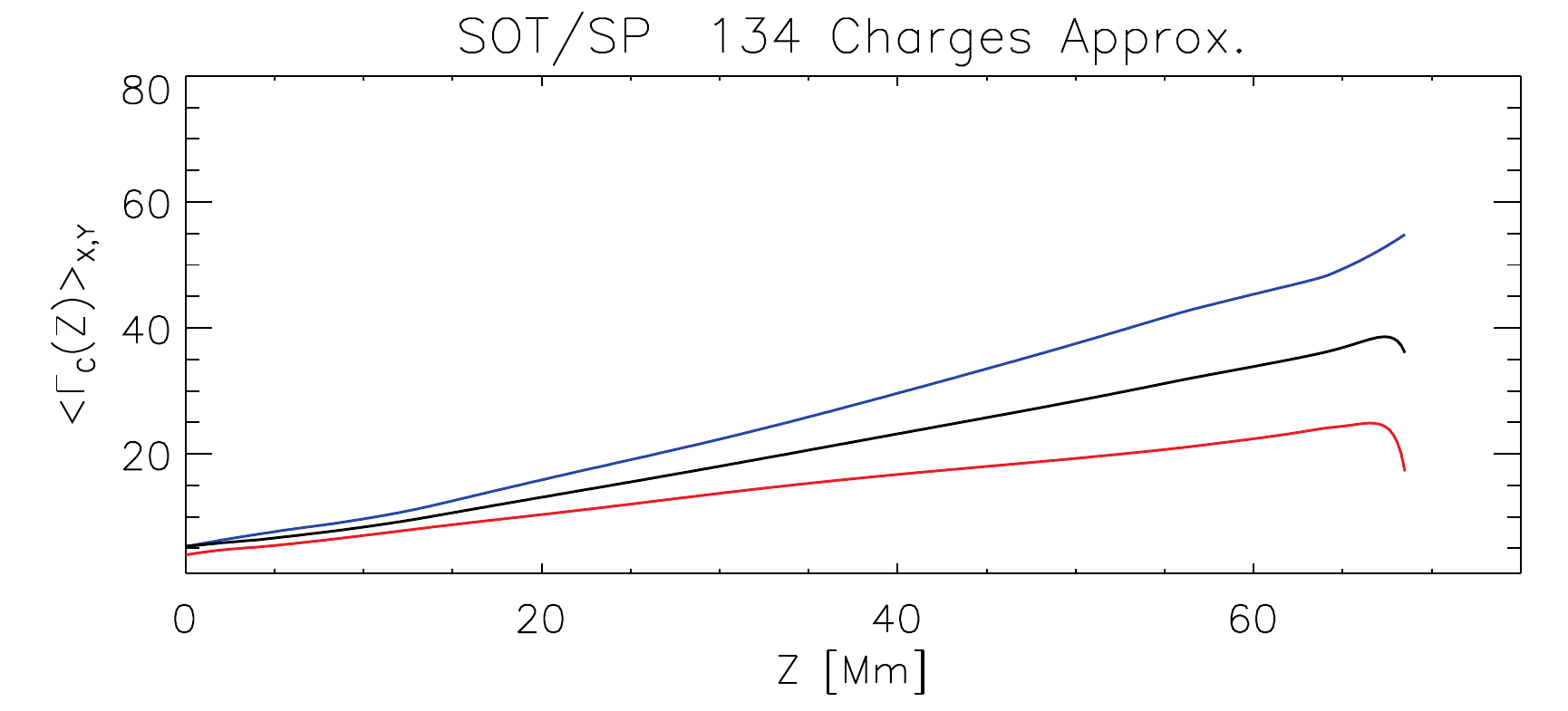}
	\caption{\textit{Top}: Histograms for the area expansion factors, produced for all grid points containing field lines ``closed'' in the computational box. \textit{Bottom}: Height-averaged values of $\left<\Gamma(Z)\right>_{X,Y}$ Green's function extrapolation (\textit{left}) and approximation by 134 submerged charges (\textit{right}).}
	\label{Fig:Gamma}
   \end{figure*}
%

\subsection{Approximation by Submerged Charges}
\label{Sect:3.2}

We also performed an approximation of the observed magnetogram by a series of submerged charges \citep[e.g.,][]{Seehafer86,Gorbachev88,Gorbachev89,Demoulin94,Longcope05}. This approximation is done to study the role of small-scale structuring of the observed SOT/SP magnetogram on the structure of the area expansion factors. We chose this approximation rather than simple rebinning of the magnetogram since the rebinned magnetogram would contain structuring down to the size of the rebinned pixel. Note that structuring of the expansion factors calculated from the \textit{SOHO}/MDI magnetogram with $\approx$2$\arcsec$ resolution was already investigated by \citet{Dudik11}. Rather, we are interested in comparison of the calculated magnetic field (Sect. \ref{Sect:3.1}) with an approximated magnetic field at the same resolution, retaining the large-scale observed flux distribution, but without the small-scale structuring.

The approximation by submerged charges allows one to calculate the potential magnetic field by using the analogy with electrostatic field of individual charges
  \begin{equation}
	\mathbf{B}_C (\mathbf{r}) = \sum_{i=1}^N C_{i} \frac{\mathbf{r} -\mathbf{R}_{C,i}}{\,\left|\mathbf{r} -\mathbf{R}_{C,i} \right|^3} \,,
	\label{Eq:Charges}
  \end{equation}
where $\mathbf{R}_{C,i}$ is the location of the individual charge and $C_{i}$ is its amplitude. Since the charges are an artifact of the method, the above expression holds only for $Z$\,$\geq$\,0 and is not meant to represent the magnetic field below the photosphere ($Z$\,=\,0).

The approximation of the magnetogram with $N$ charges is done in the following manner. First, the observed magnetogram is rebinned by a factor of 4. This is done to remove the fine-structuring while keeping the major magnetic polarities, and thus help guide the initial placement of magnetic charges. The depth of the charges $Z_C$\,=\,const. is then estimated using the rebinned magnetogram. This estimate utilizes fact that the resulting horizontal extent of a polarity in the approximated magnetogram is comparable to the depth where charges are placed \citep{Demoulin94}. Then, the locations $X_{C,i}$ and $Y_{C,i}$ of the charges, 1\,$\leq$\,$i$\,$\leq$\,$N$, are determined manually to approximate the distribution of magnetic polarities in the rebinned magnetogram. The charges, $C_i$ are then determined by least-square fitting of the resulting flux distribution with the flux distribution in the original magnetogram. In the next iteration, the $X_{C,i}$, $Y_{C,i}$ are varied, and then the new values of $C_i$ are determined. The whole procedure is repeated for a set of $Z_C$, and the best-match to the observed magnetogram is determined based on minimum difference between the observed and approximated magnetograms. We note that there is no single ``best-match'' approximation to the observed magnetogram, as both the number of charges and their depth must be chosen manually. 

Since the resolution of the SOT/SP magnetogram is high, many individual charges are needed to produce a reasonable match to the observed distribution of magnetic flux. Upon many trial-and-error, we determined that a distribution of 134 individual charges corresponding to $Z_C$\,=\,3.7\,Mm (17 px) below the photosphere gives a sufficient approximation of the observed flux distribution (Fig. \ref{Fig:SOT}, \textit{top}). We note that a much smaller depth $Z_C$ might lead to a closer representation of some of the fine-structuring of the observed magnetogram. However, the number of submerged charges would then have to be correspondingly larger, making the approximation impractical. We also note that \citet{Aschwanden14} found that $\approx$10$^2$ charges is typically sufficient to approximate an active region.

We also note that we chose the submerged charge method instead of simply decreasing the resolution of the magnetogram, since the structure of the expansion factor for a \textit{SOHO}/MDI magnetogram \citep{Scherrer95} with a $\approx$2$\arcsec$ resolution has already been done by \citet{Dudik11}.

Finally, we note that the magnetic field obtained using Eq. (\ref{Eq:Charges}) is very similar to that obtained using the direct extrapolation by the Green's function method (Fig. \ref{Fig:SOT}, \textit{middle}). 

   \begin{figure*}[!ht]
	\centering
	\includegraphics[width=8.8cm]{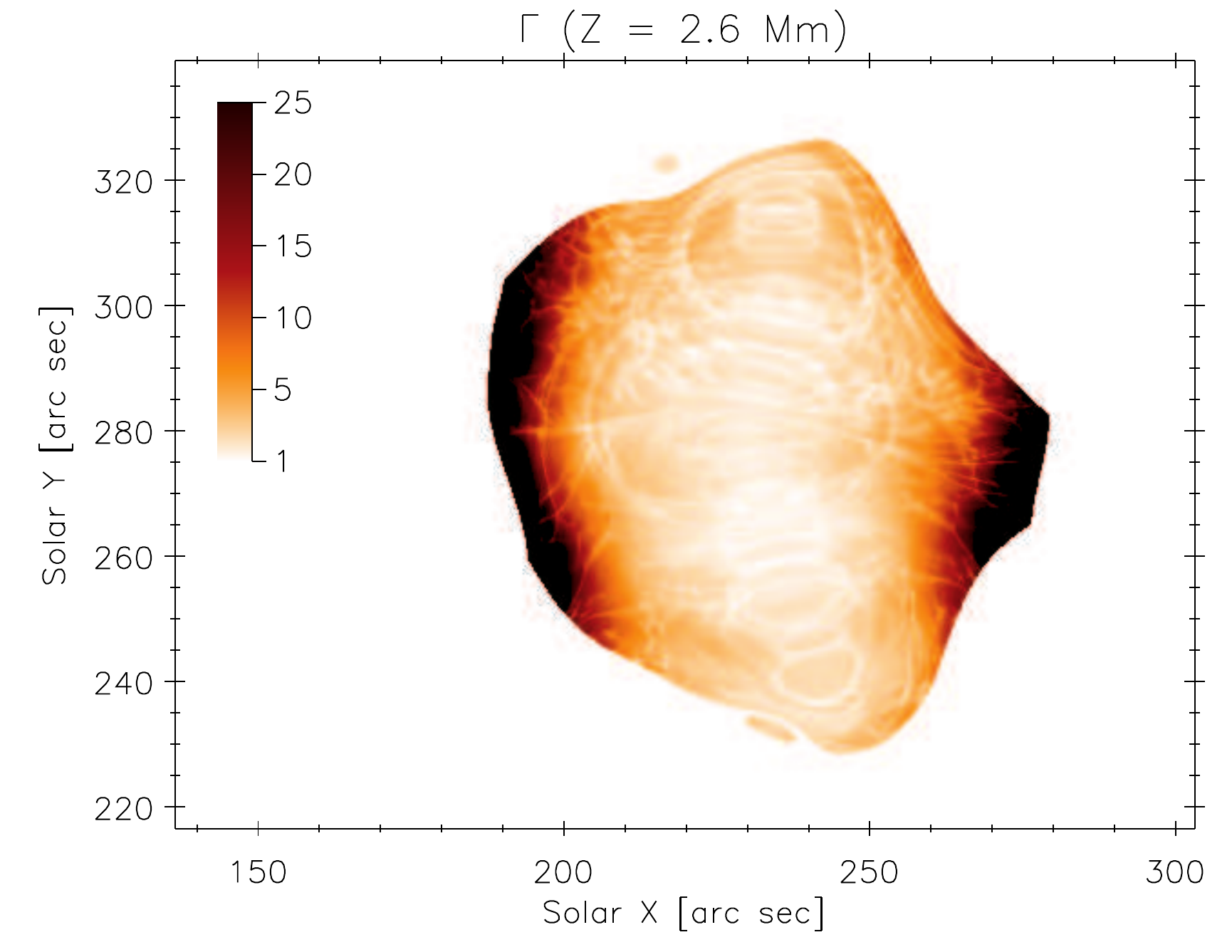}
	\includegraphics[width=8.8cm]{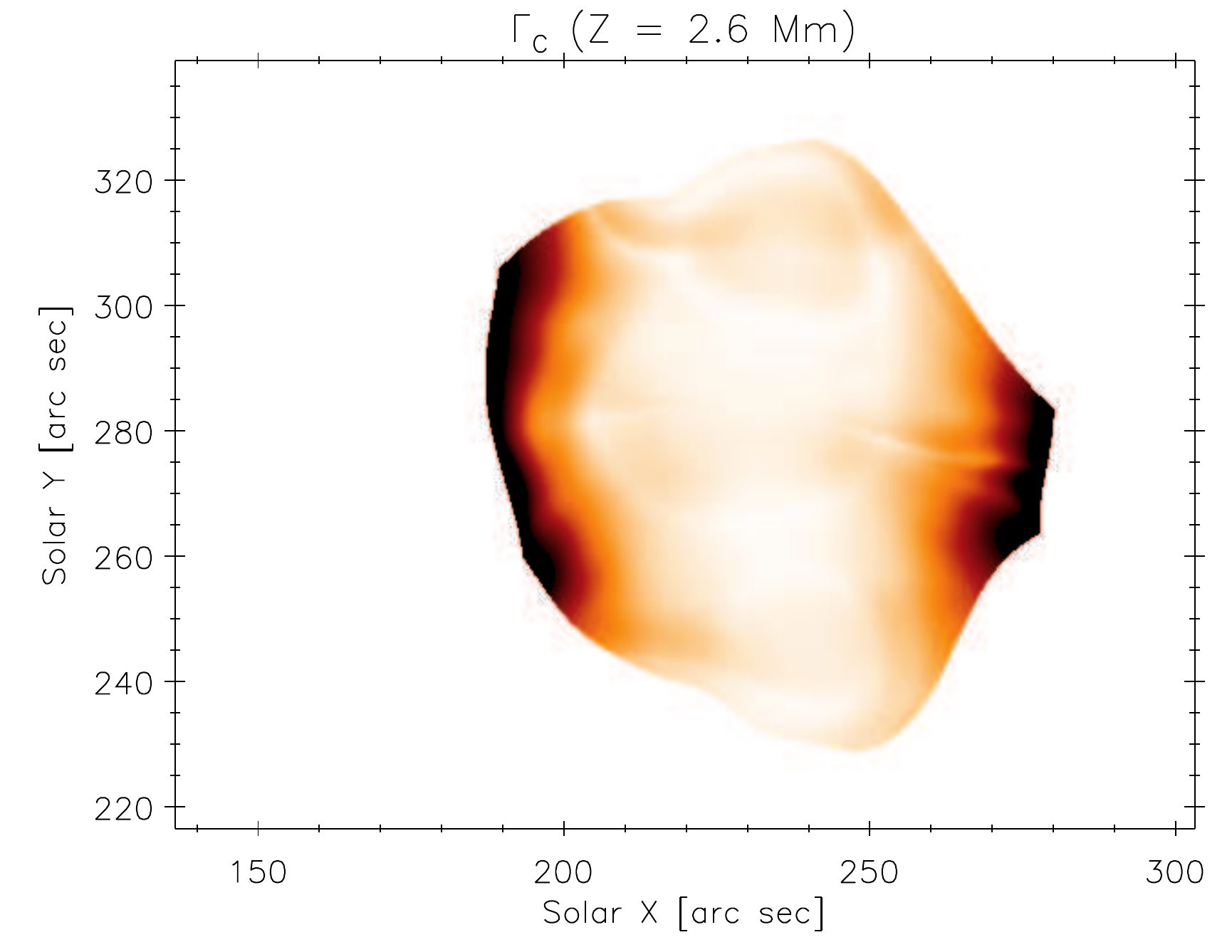}
	\includegraphics[width=8.8cm]{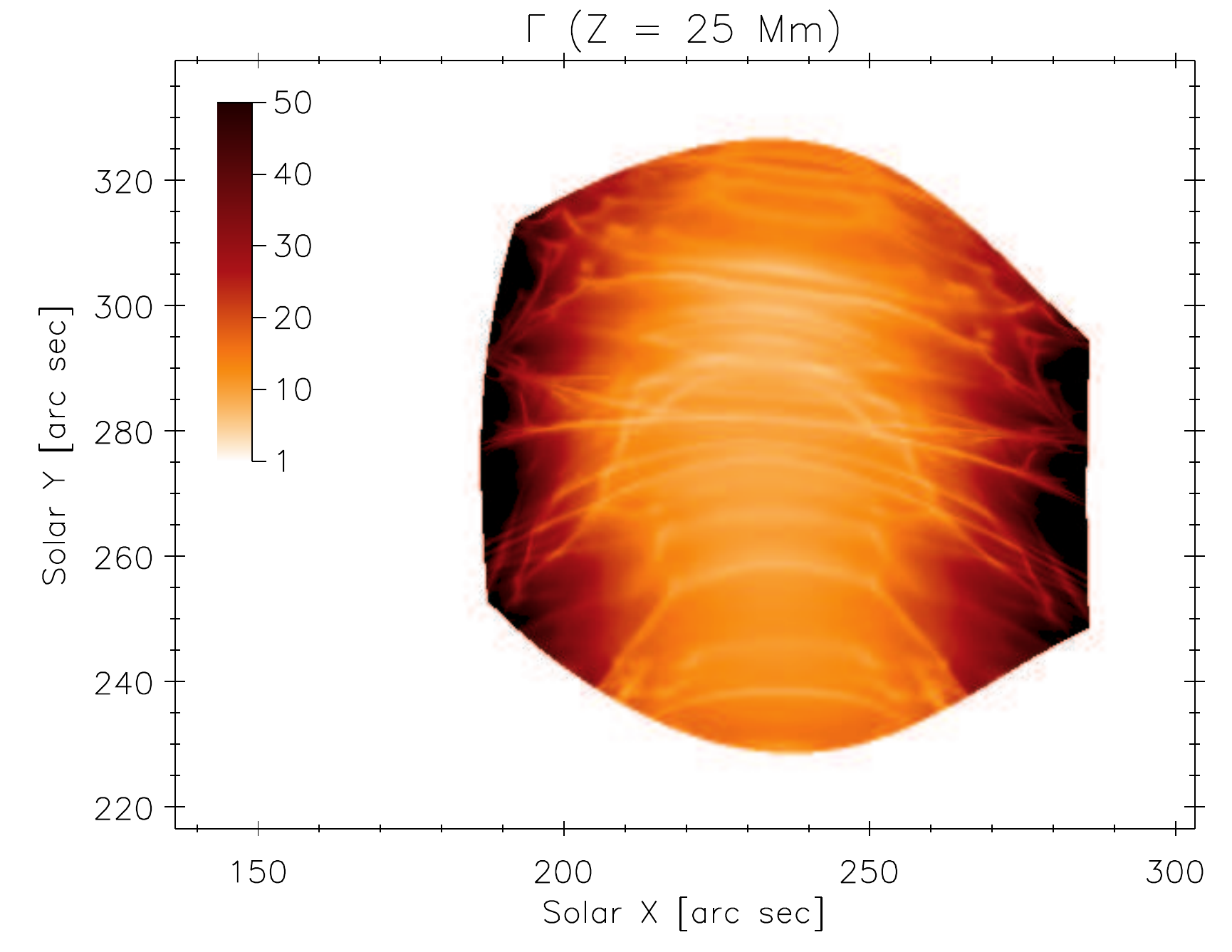}
	\includegraphics[width=8.8cm]{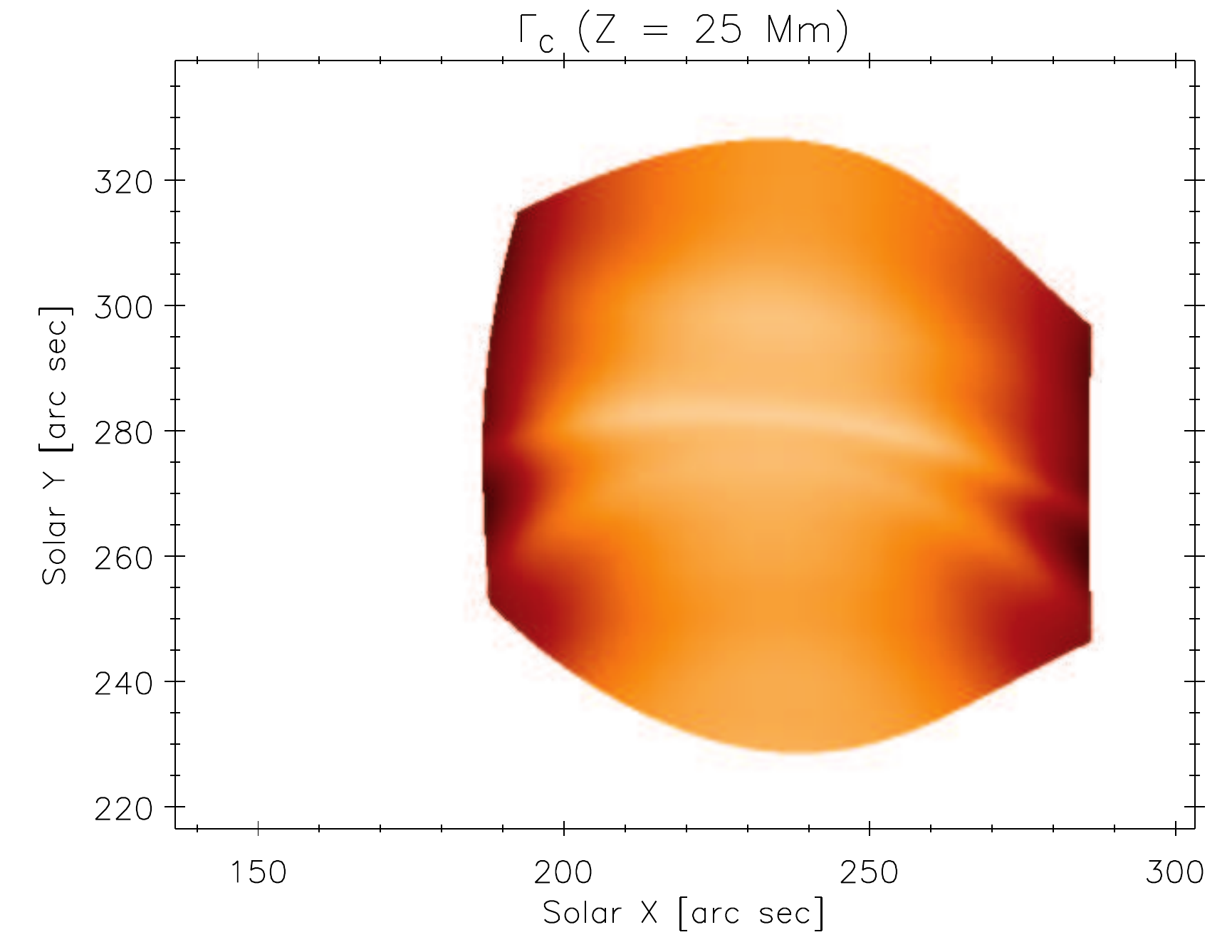}
	\includegraphics[width=8.8cm]{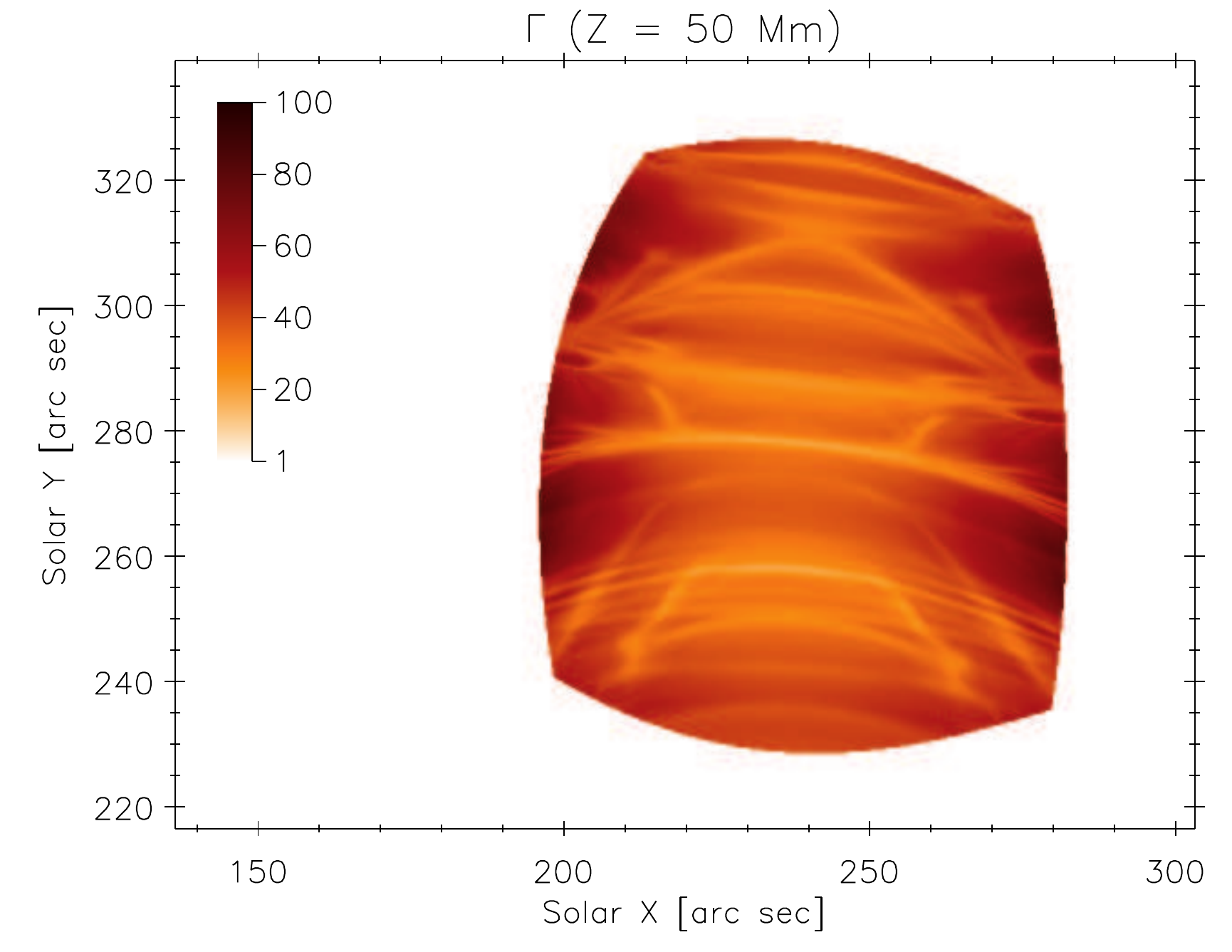}
	\includegraphics[width=8.8cm]{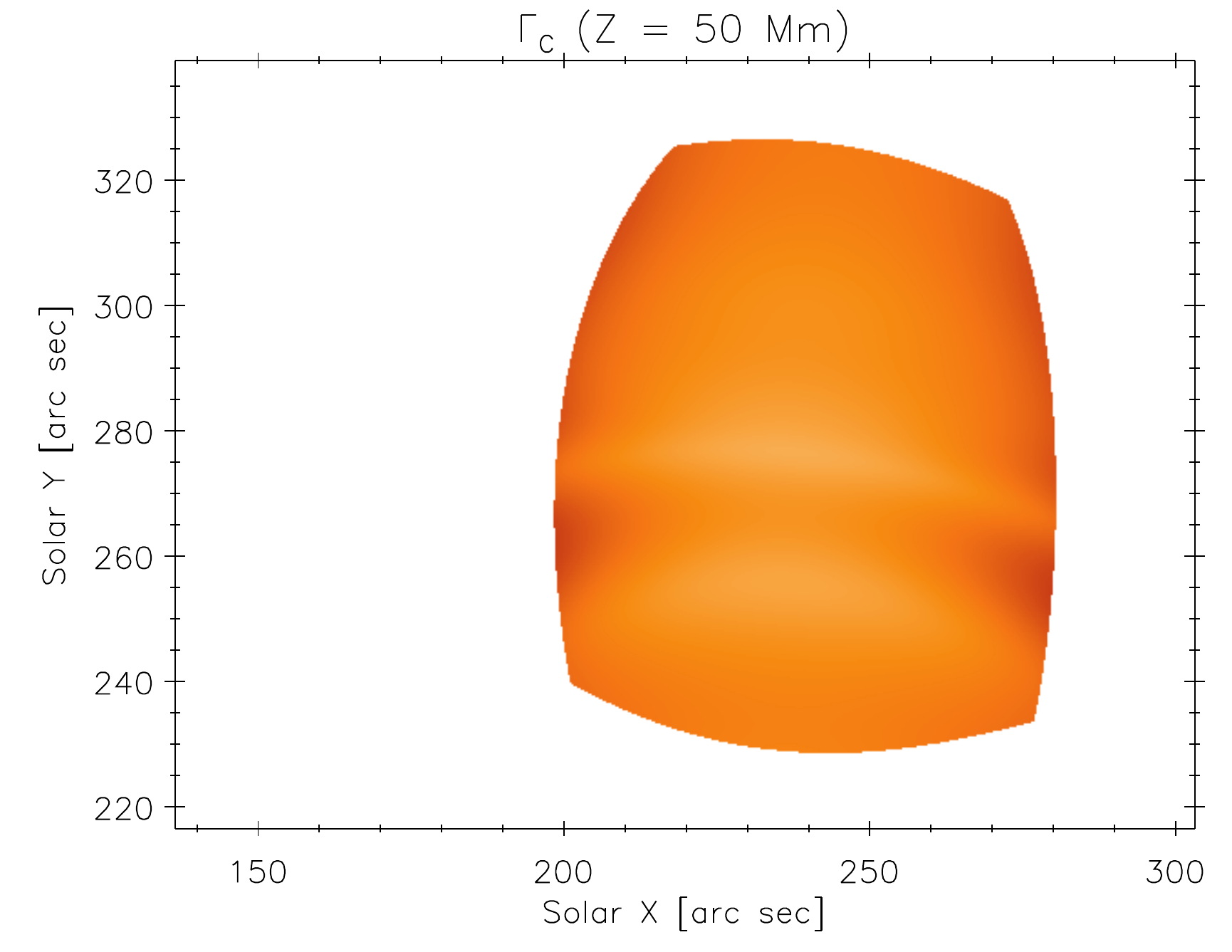}
	\caption{Horizontal cuts through the spatial distribution of $\Gamma$ (\textit{left}) and $\Gamma_C$ (\textit{right}) at heights of $Z$\,=\,2.6\,Mm (\textit{top}), 25\,Mm (\textit{middle}) and 50\,Mm (\textit{bottom}). }
	\label{Fig:Gamma_cutZ}
   \end{figure*}
%
%
   \begin{figure*}[!ht]
	\centering
	\includegraphics[width=8.8cm]{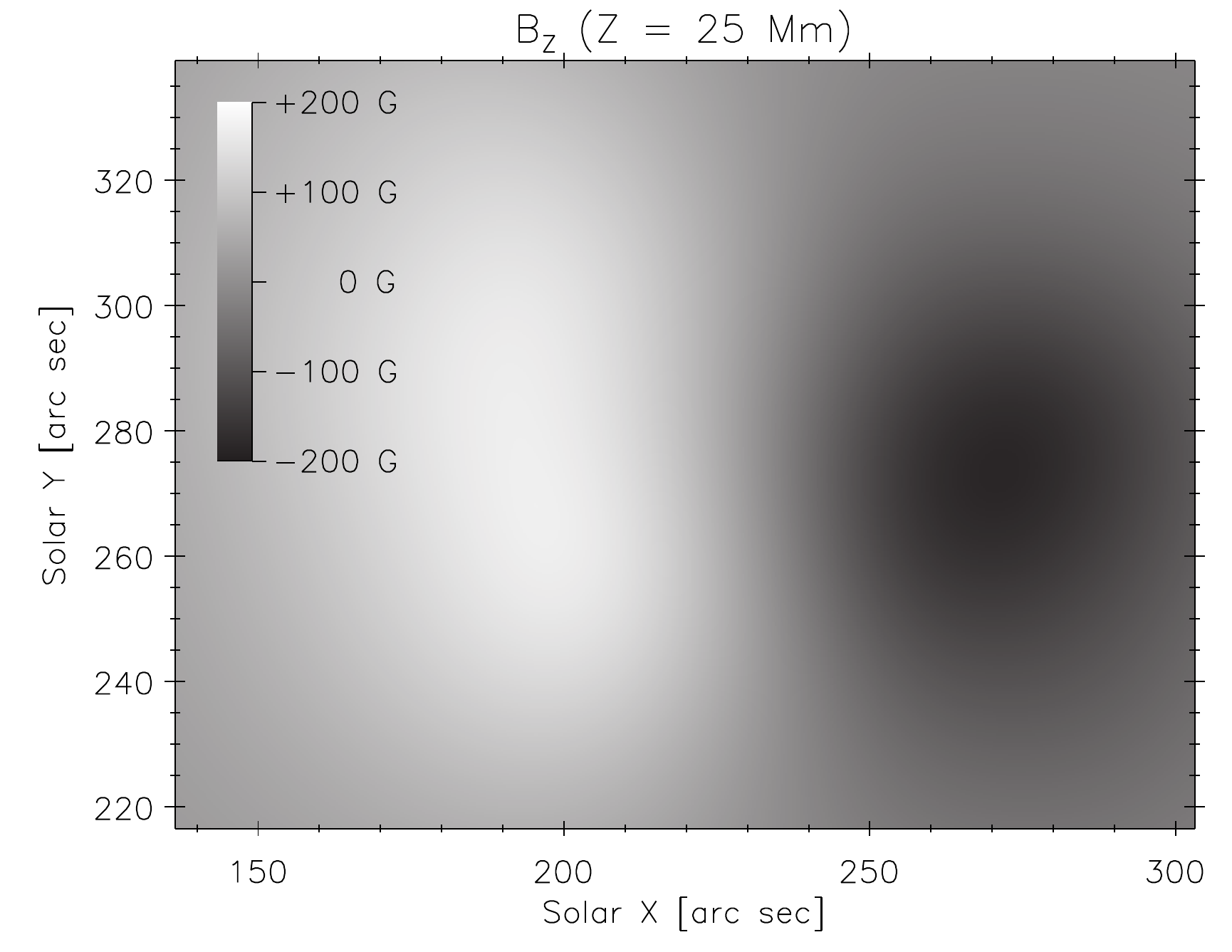}
	\includegraphics[width=8.8cm]{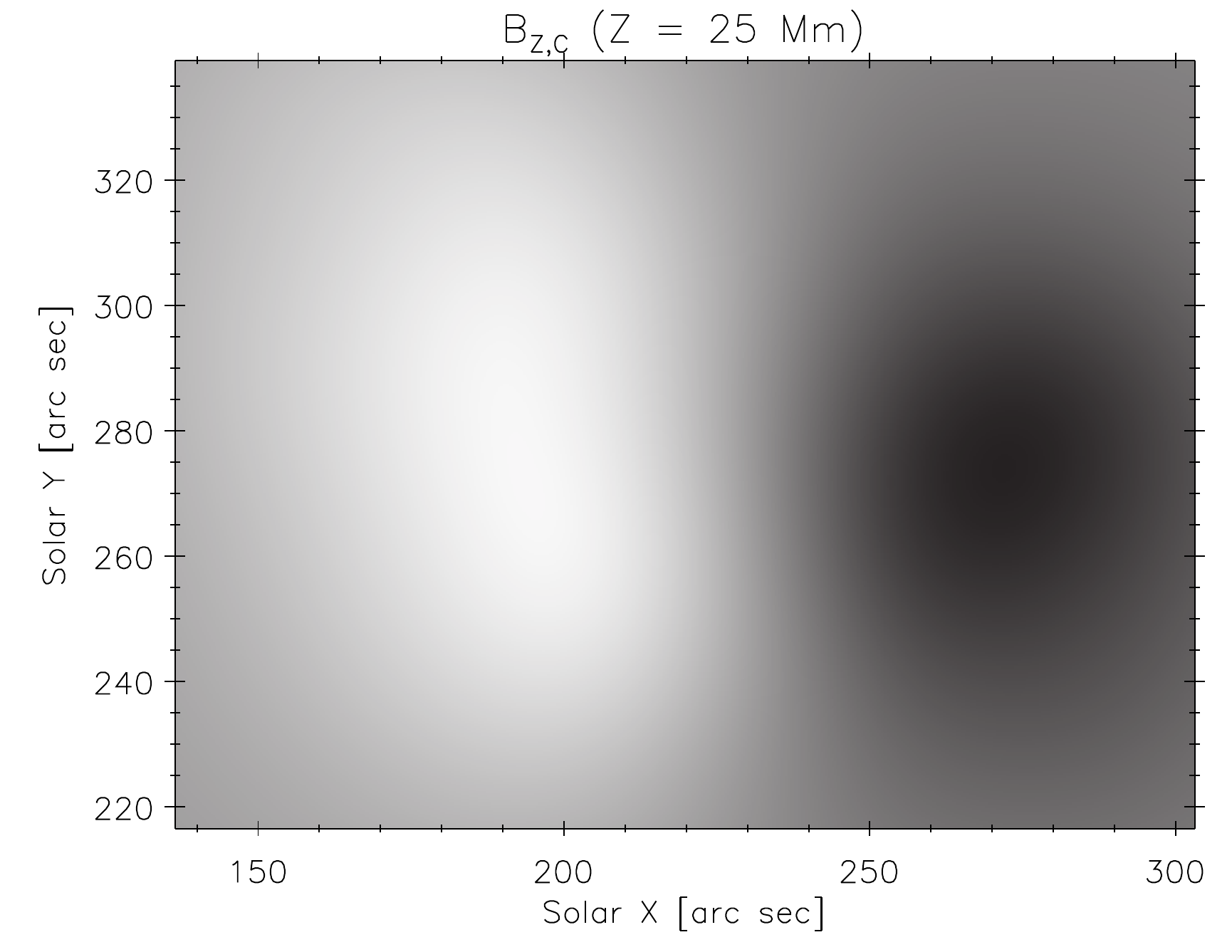}
	\caption{Distribution of $B_Z$ (\textit{left}) and $B_{Z,C}$ (\textit{right}) at $Z$\,=\,25\,Mm.}
	\label{Fig:BZ_cutZ}
   \end{figure*}

\subsection{Calculation of the Area Expansion}
\label{Sect:3.3}

The area expansion factor is calculated for both the directly extrapolated magnetic field, as well as the magnetic field obtained from the submerged charges approximation. The calculation method is that of \citet{Dudik11}. A field line passing through each grid point $\mathbf{r}$\,=\,[$x$,\,$y$\,$z$] is traced using the fourth-order Runge-Kutta method. This line is identified with a loop strand. Note that, in accordance with the literature, we use the term ``loop'' for an observed structure, while the term ``strand'' is used for a fundamental, independently heated, thermally isolated, and possibly unresolved magnetic structure. The area expansion factor $\Gamma(\mathbf{r})$ of the traced strand is calculated as the ratio of the strand cross-section at the apex to the cross-section at the photosphere. Since magnetic flux is conserved, the $\Gamma(\mathbf{r})$ can be defined as \citep[see also][]{Asgari13}
  \begin{equation}
	\Gamma(\mathbf{r}) = \frac{\Gamma_{+}(\mathbf{r}) + \Gamma_{-}(\mathbf{r})}{2} =  \frac{B_{+}(\mathbf{r})+ B_{-}(\mathbf{r})}{2B_{\mathrm{apex}}(\mathbf{r})}\,,
	\label{Eq:Gamma}
  \end{equation}
where the $B_{+}(\mathbf{r})$ and $B_{-}(\mathbf{r})$ is the magnetic field at the positive and negative photospheric footpoints of the field line passing through $\mathbf{r}$. Similarly, $B_{\mathrm{apex}}(\mathbf{r})$ is the magnetic field induction at the apex (top) of the field line, i.e., at the location where $B_Z$\,=\,0. The $\Gamma(\mathbf{r})$ is calculated only for a set of grid points $(\mathbf{r})$ that are located on a magnetic field line ``closed'' within the computational box. This is because for the field lines ``open'' within the box, one or more of the quantities at the right-hand side of Eq. (\ref{Eq:Gamma}) cannot be properly defined.

The area expansion factor $\Gamma(\mathbf{r})$ in Eq. (\ref{Eq:Gamma}) is defined as a global quantity for a field line passing through $\mathbf{r}$. That is, all points $s$ along the same field line will be assigned the same value of $\Gamma$. This is done in order to identify whether there are field lines in close neighborhood having different $\Gamma$. The local area expansion at the point $s(\mathbf{r})$ along the field line can be easily obtained from the $B(s)$ profile along the field line. The $\Gamma(\mathbf{r})$, as defined in Eq. (\ref{Eq:Gamma}), is one of the main factors determining the total volume of the loop strand, with the other ones being the length $L$ and the (arbitrarily chosen) size of the cross-section at some pre-defined point along the strand (see Appendix \ref{Appendix:A}, Eq. \ref{Eq:V_total}).

Note that the proposed coronal heating mechanisms can be parametrised as a function of the magnetic field \citep{Mandrini00,Lundquist08}. In in Eq. (\ref{Eq:Gamma}), the averaging over the magnetic field at both photospheric footpoints, $B_{+}$ and $B_{-}$, is done to reflect the total heat input into the coronal loop strand rather than a heating input into only one portion of the strand. Coronal loops are rarely geometrically symmetric. Similarly, the $B_{+}$ and $B_{-}$ are not the same for a given strand. Lack of symmetry implies that the temperature maximum is located away from the apex \citep[e.g.,][]{Klimchuk10}. This means that the two un-equal halves of the strand are not thermally isolated, since thermal conduction does not vanish at the apex. Furthermore, evolving loops can exhibit flows and ethalpy fluxes \citep[e.g.,][]{DelZanna08,Bradshaw08,Bradshaw10a,Bradshaw10b,Marsch08,Tripathi09,Klimchuk10,Warren11,Young12,Scott12,Tripathi12,Winebarger13,Mikic13,Taroyan14}, making the total heating of the strand important. This is reflected by the averaging of $B_{+}$ and $B_{-}$ in Eq. (\ref{Eq:Gamma}).

%
   \begin{figure}[!ht]
	\centering
	\includegraphics[width=8.8cm]{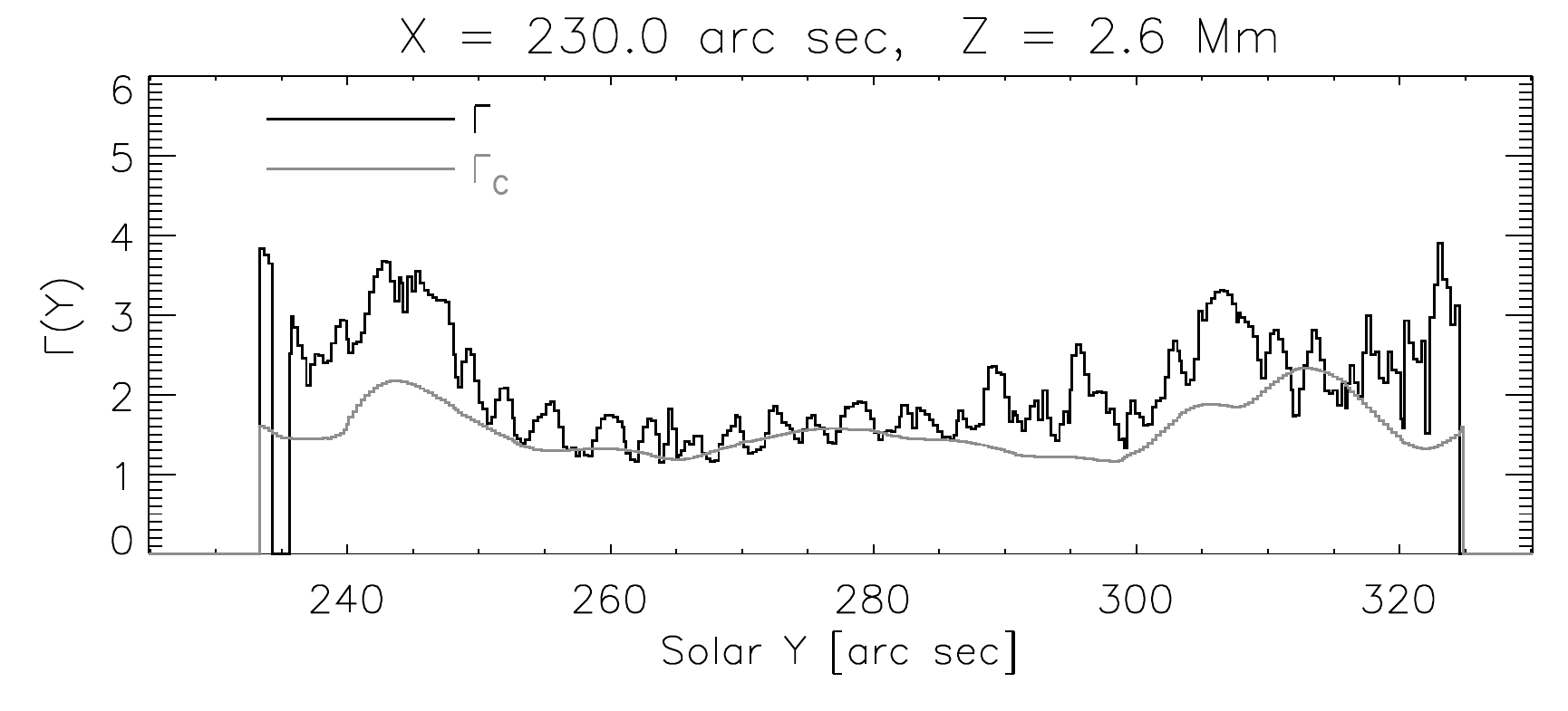}
	\includegraphics[width=8.8cm]{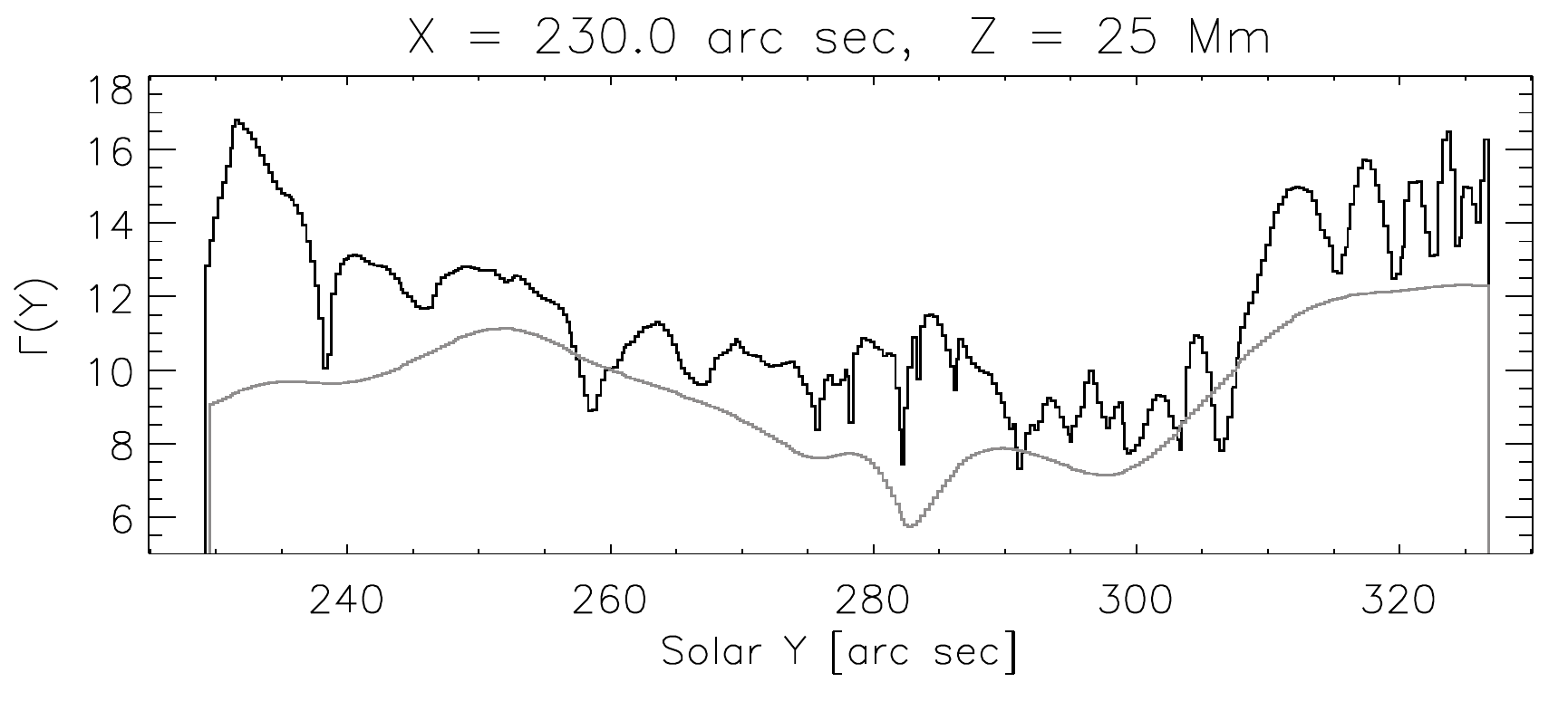}
	\includegraphics[width=8.8cm]{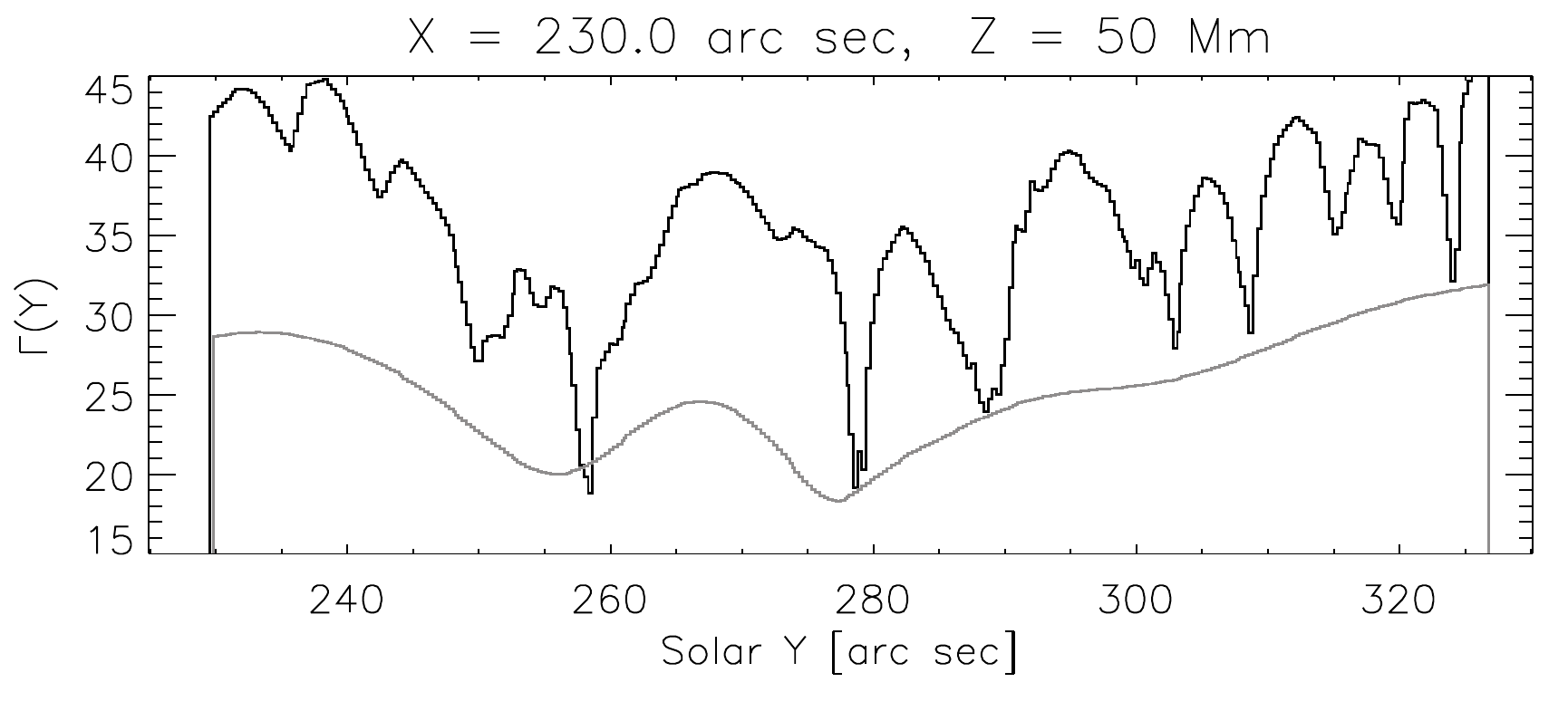}
	\caption{Variation of the $\Gamma$ (black) and $\Gamma_C$ (grey) at $X$\,=\,230$\arcsec$ and three different heights of $Z$\,=\,2.6\,Mm, 25\,Mm, and 50\,Mm, corresponding to those shown in Fig. \ref{Fig:Gamma_cutZ}.}
	\label{Fig:Gamma_cutXZ}
   \end{figure}
%
%
   \begin{figure*}[!ht]
	\centering
	\includegraphics[width=8.8cm]{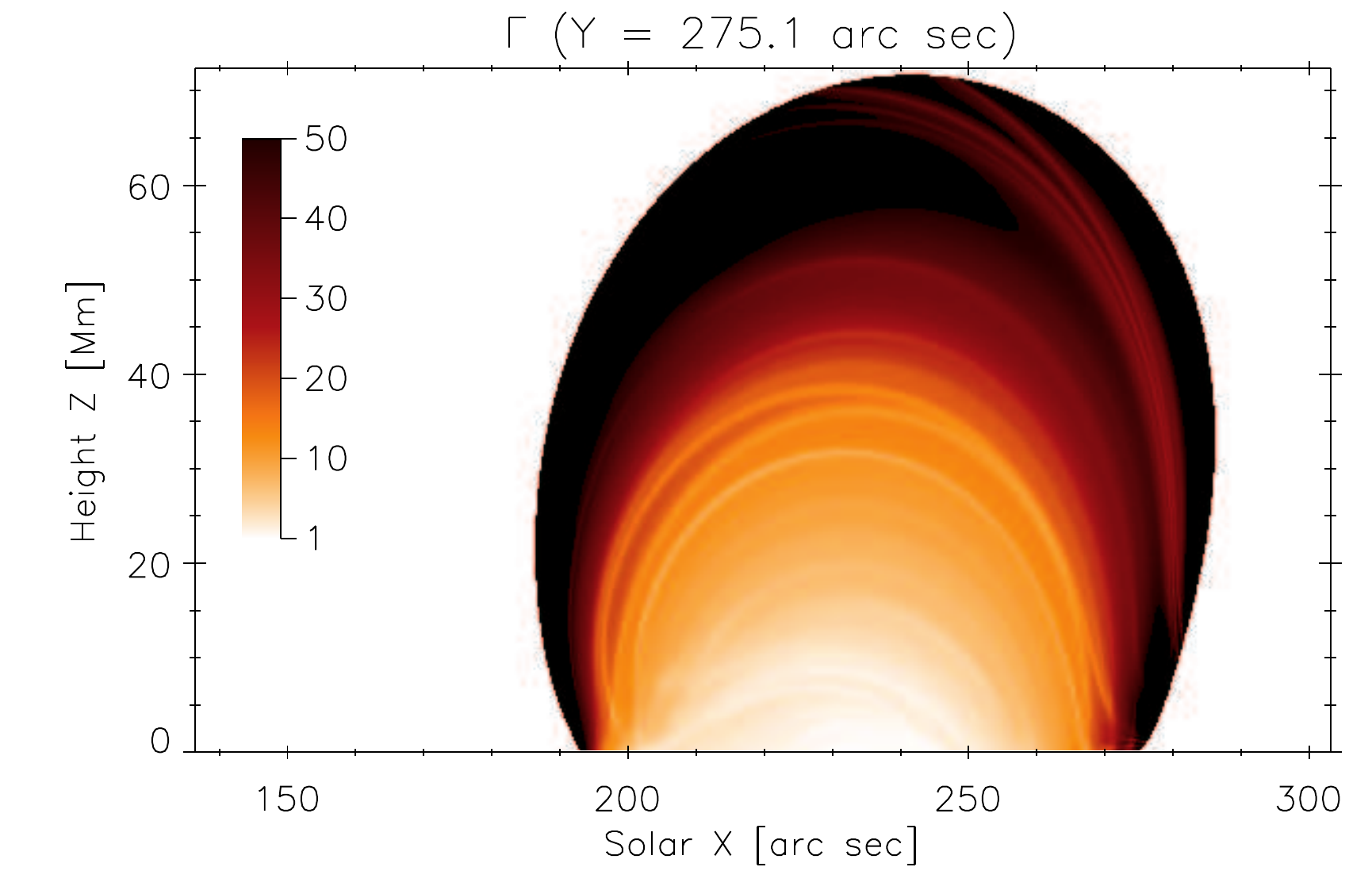}
	\includegraphics[width=8.8cm]{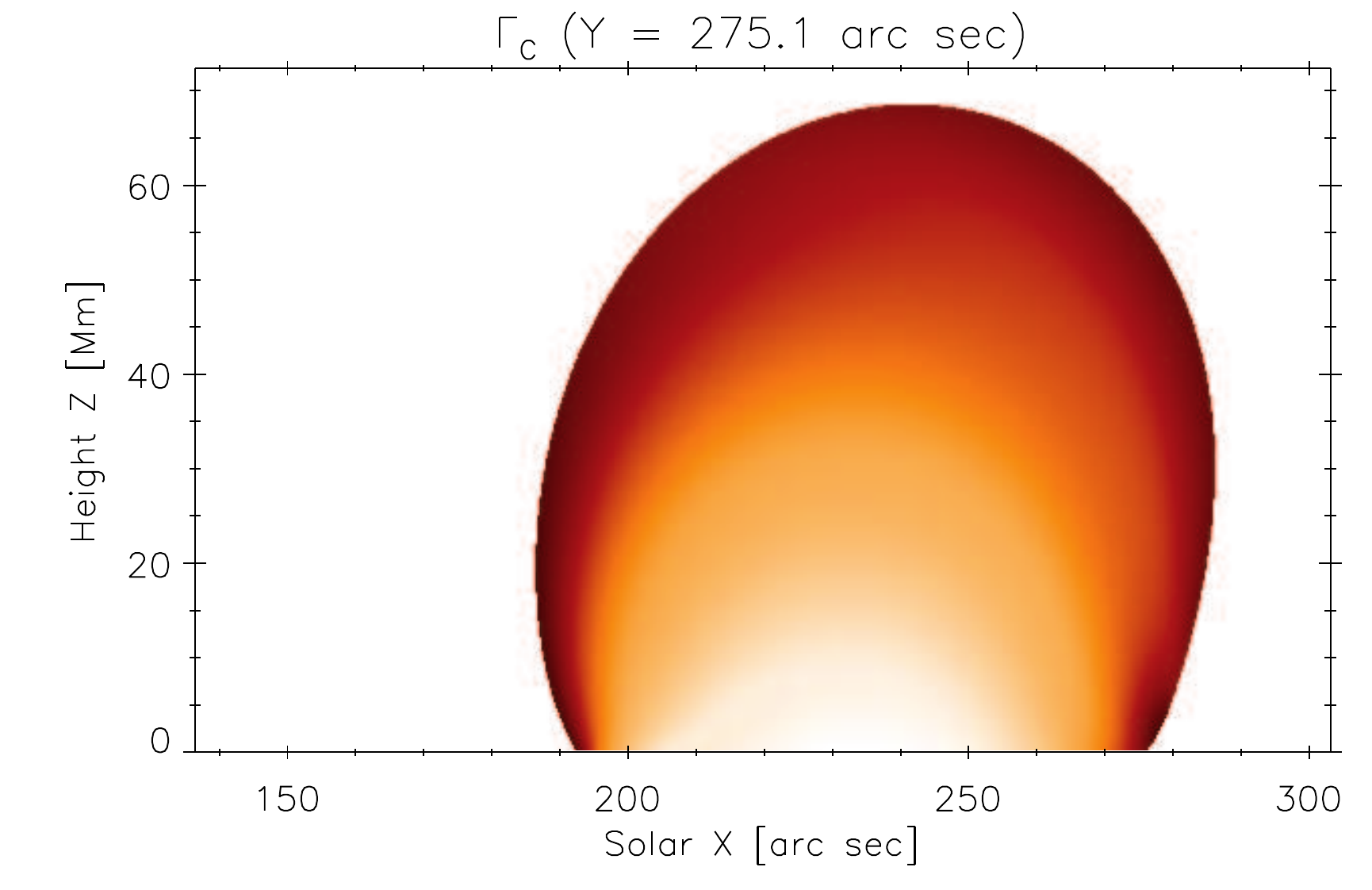}
	\includegraphics[width=8.8cm]{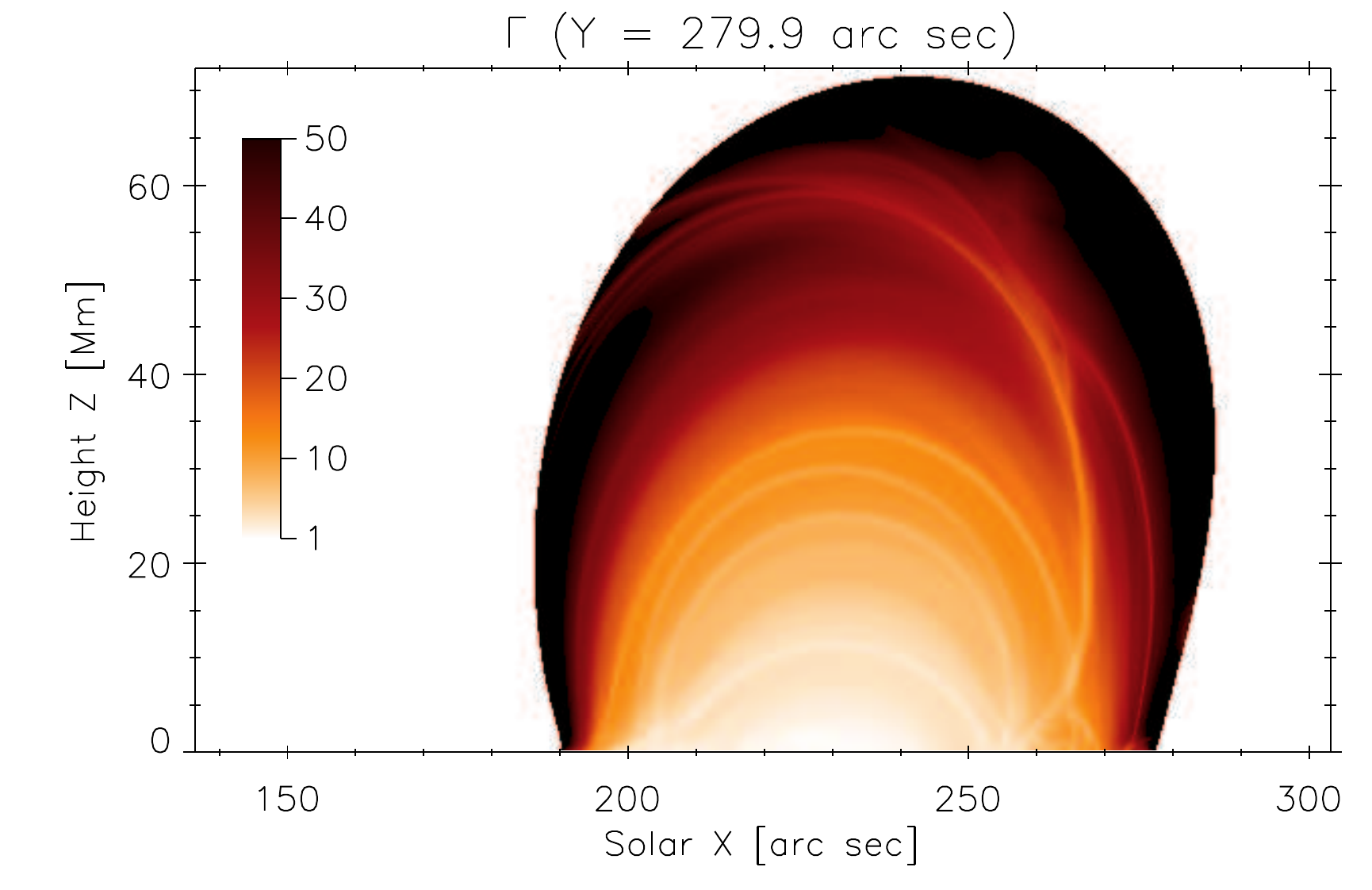}
	\includegraphics[width=8.8cm]{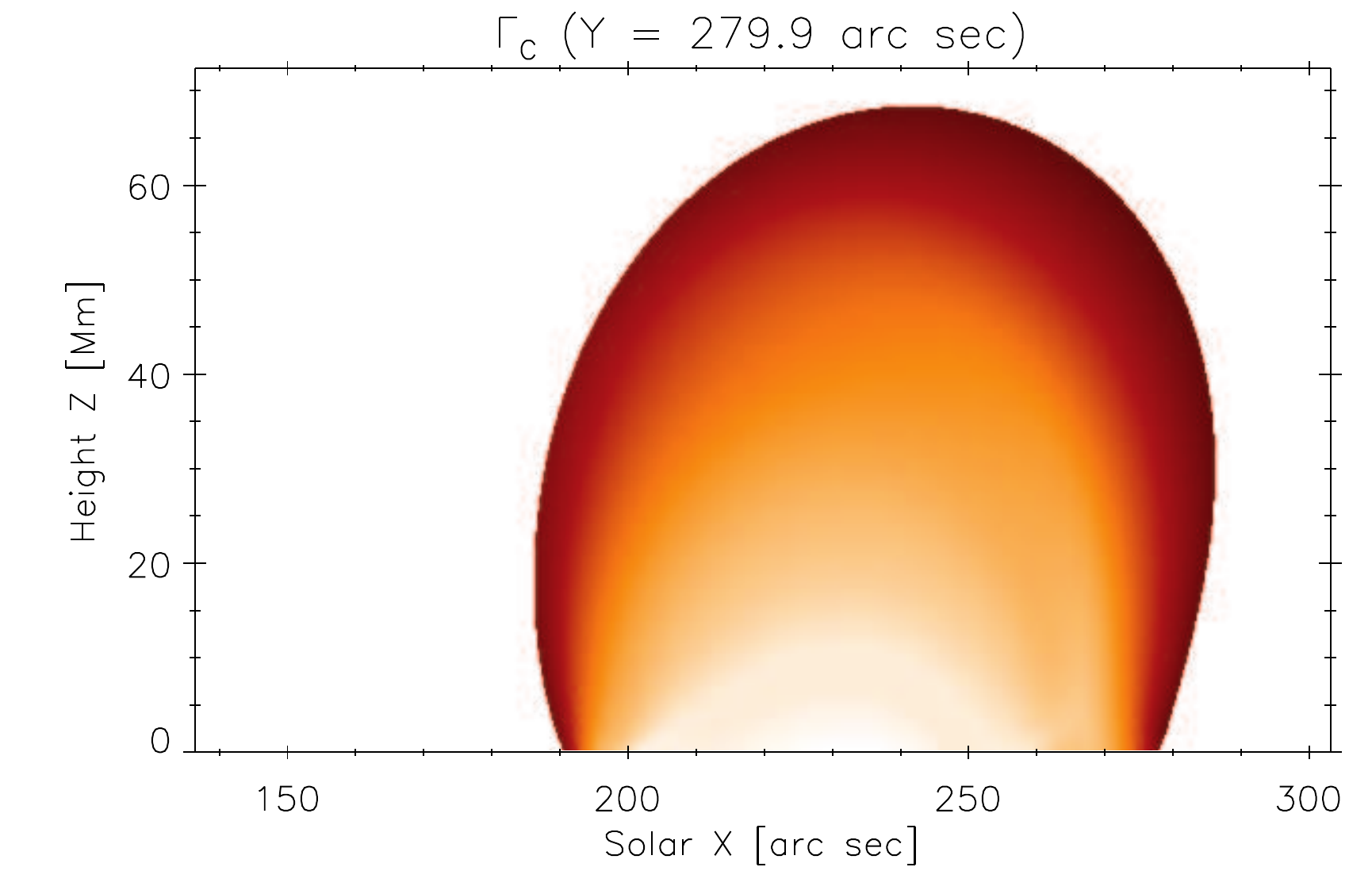}
	\includegraphics[width=8.8cm]{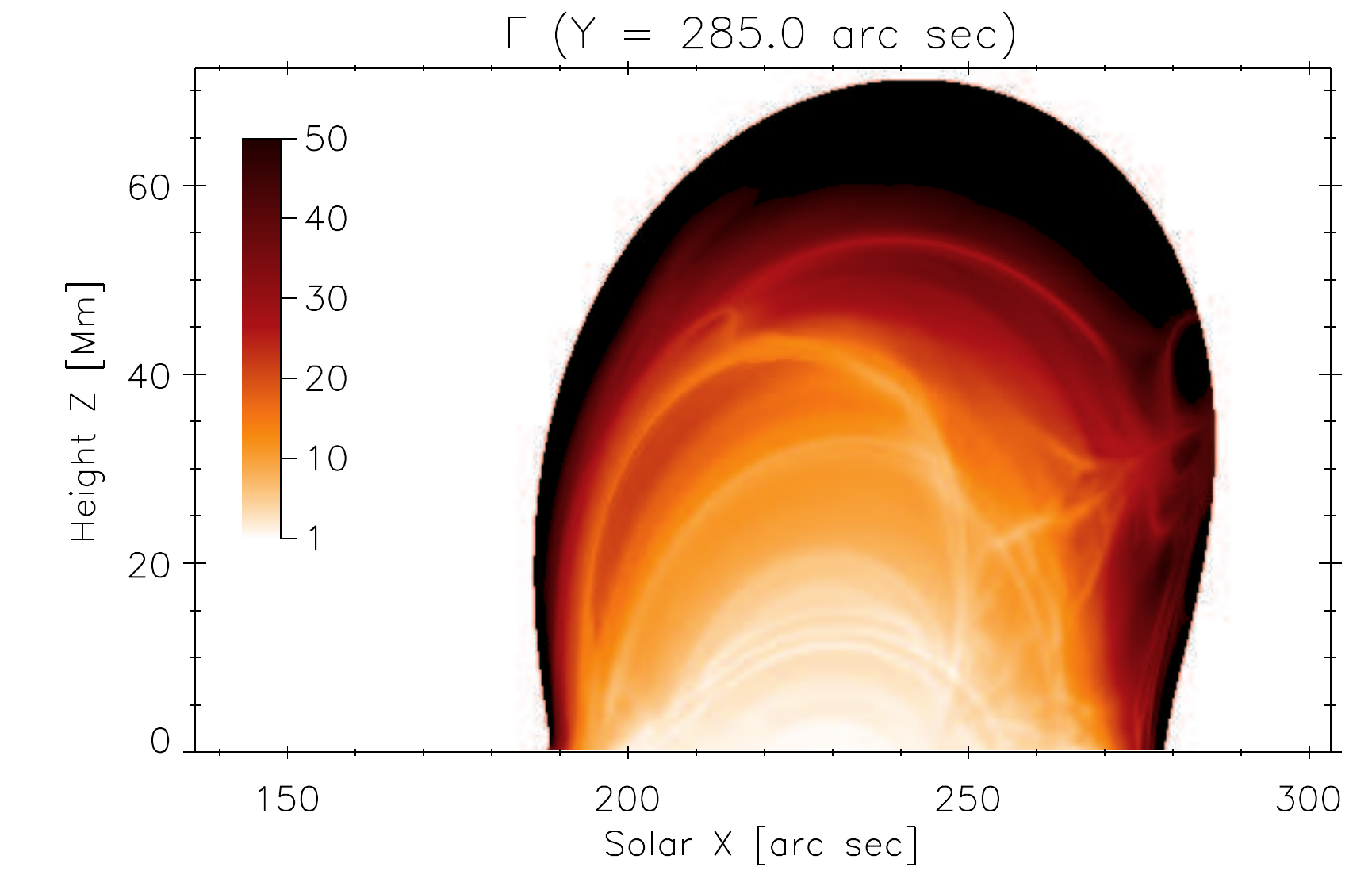}
	\includegraphics[width=8.8cm]{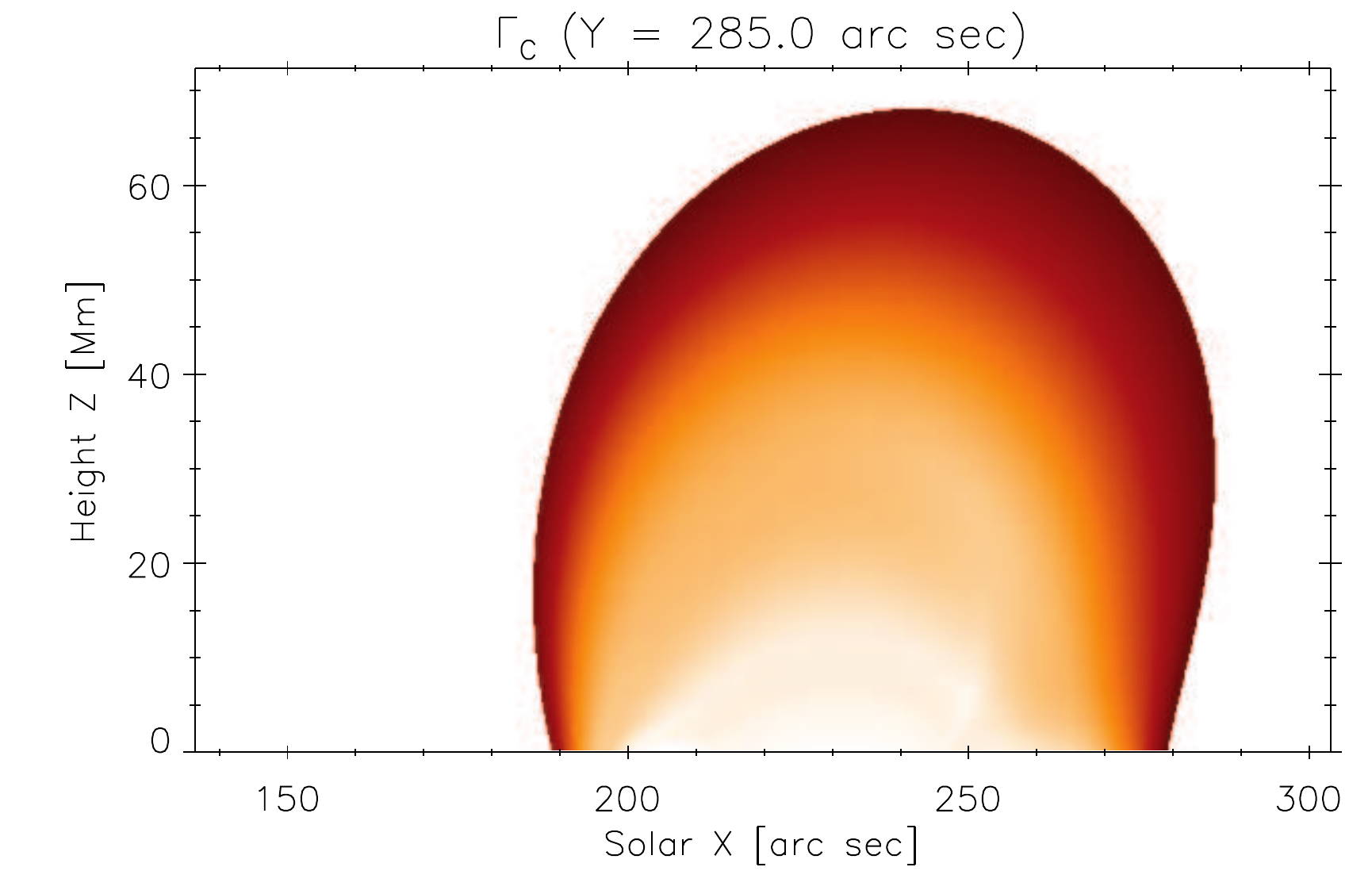}
	\caption{Vertical cuts through the spatial distribution of $\Gamma$ (\textit{left}) and $\Gamma_C$ (\textit{right}) at $Y$\,=\,275.1$\arcsec$ (\textit{top}), 279.9$\arcsec$ (\textit{middle}) and 285.0$\arcsec$ (\textit{bottom}).}
	\label{Fig:Gamma_cutY}
   \end{figure*}
%
   \begin{figure*}[!ht]
	\centering
	\includegraphics[width=6.45cm]{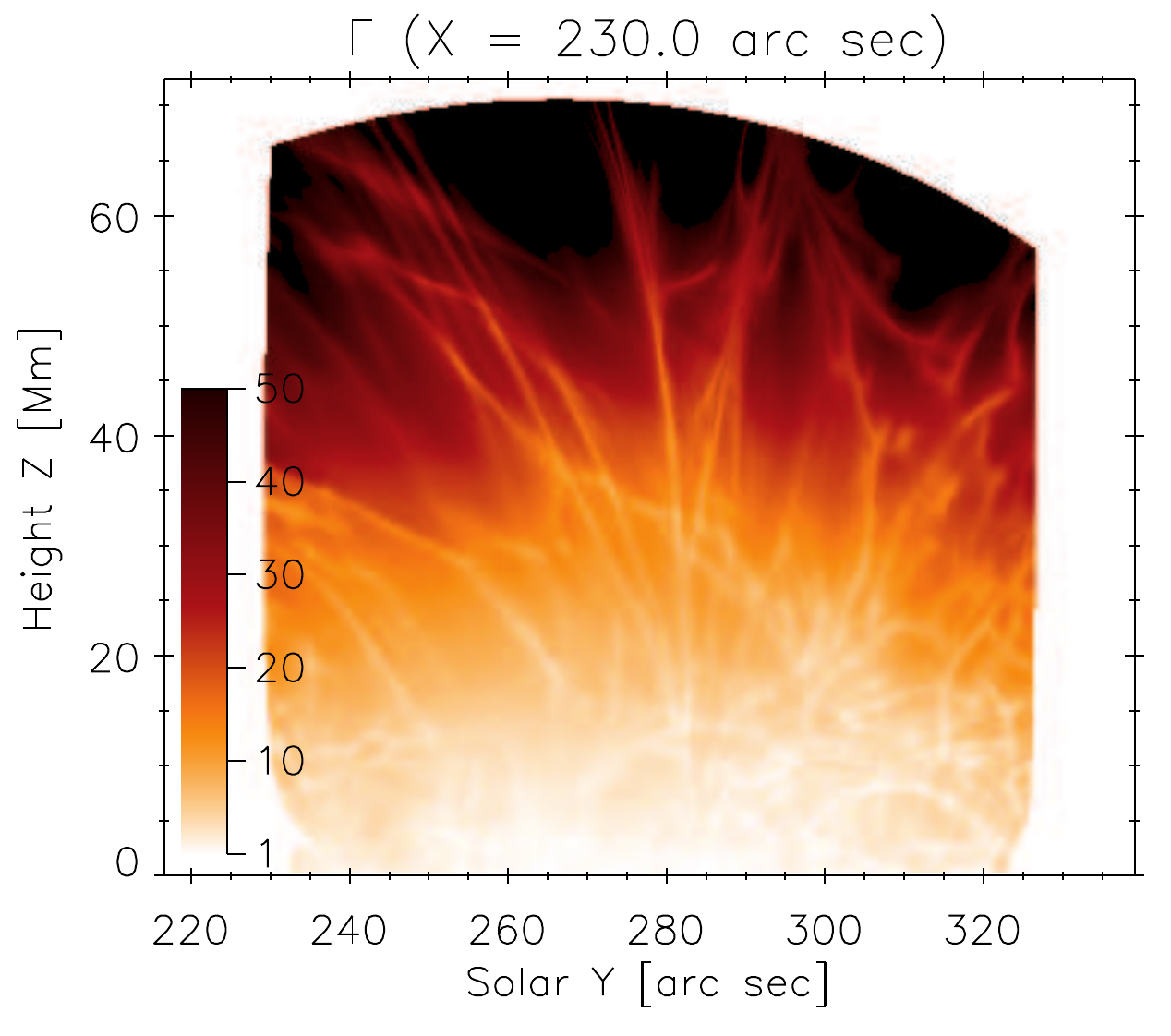}
	\includegraphics[width=6.45cm]{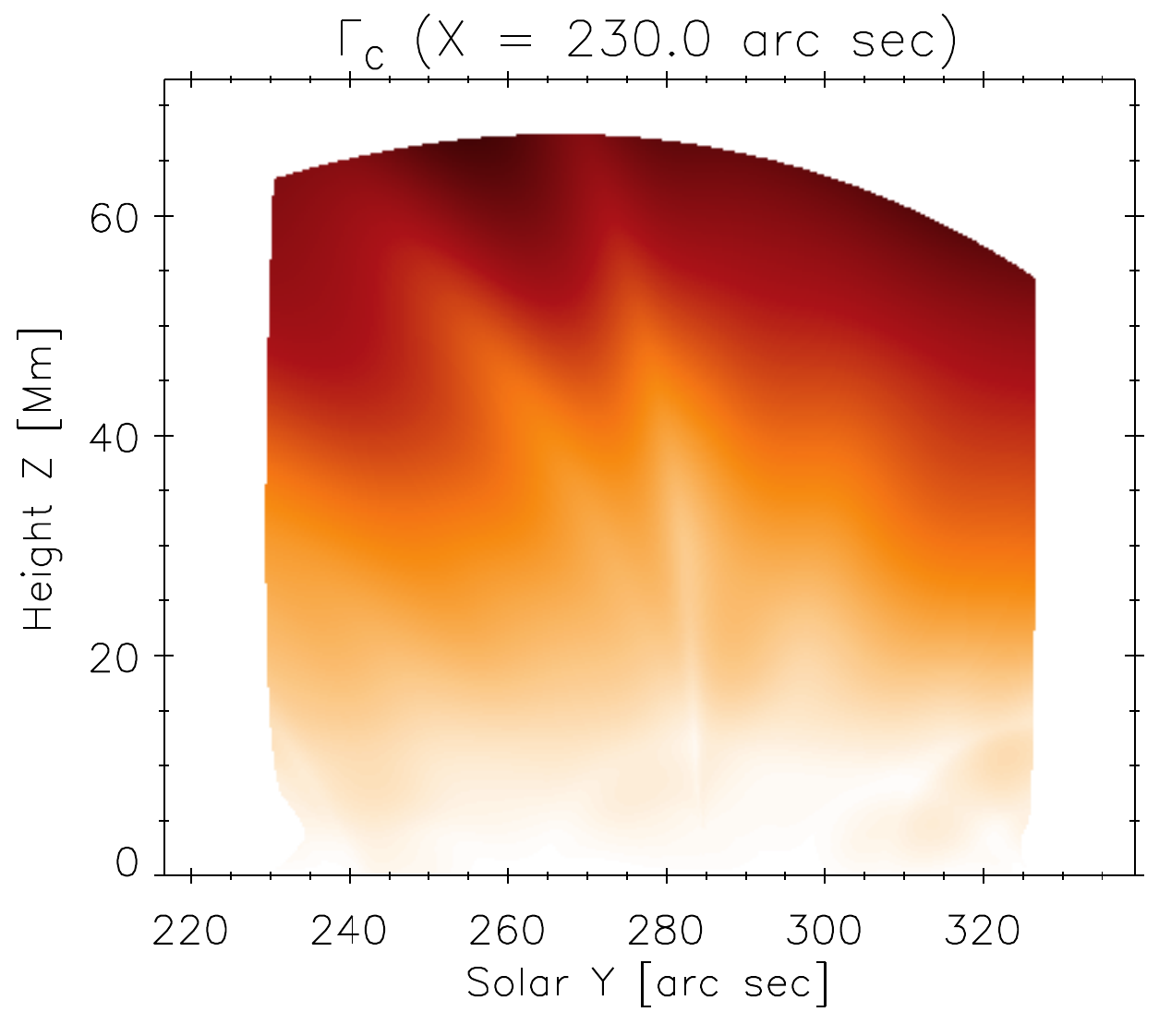}
	\caption{Vertical cuts through the spatial distribution of $\Gamma$ (\textit{left}) and $\Gamma_C$ (\textit{right}) at $X$\,=\,230$\arcsec$.}
	\label{Fig:Gamma_cutX}
   \end{figure*}
%
%
   \begin{figure*}[!ht]
	\centering
	\includegraphics[width=8.8cm]{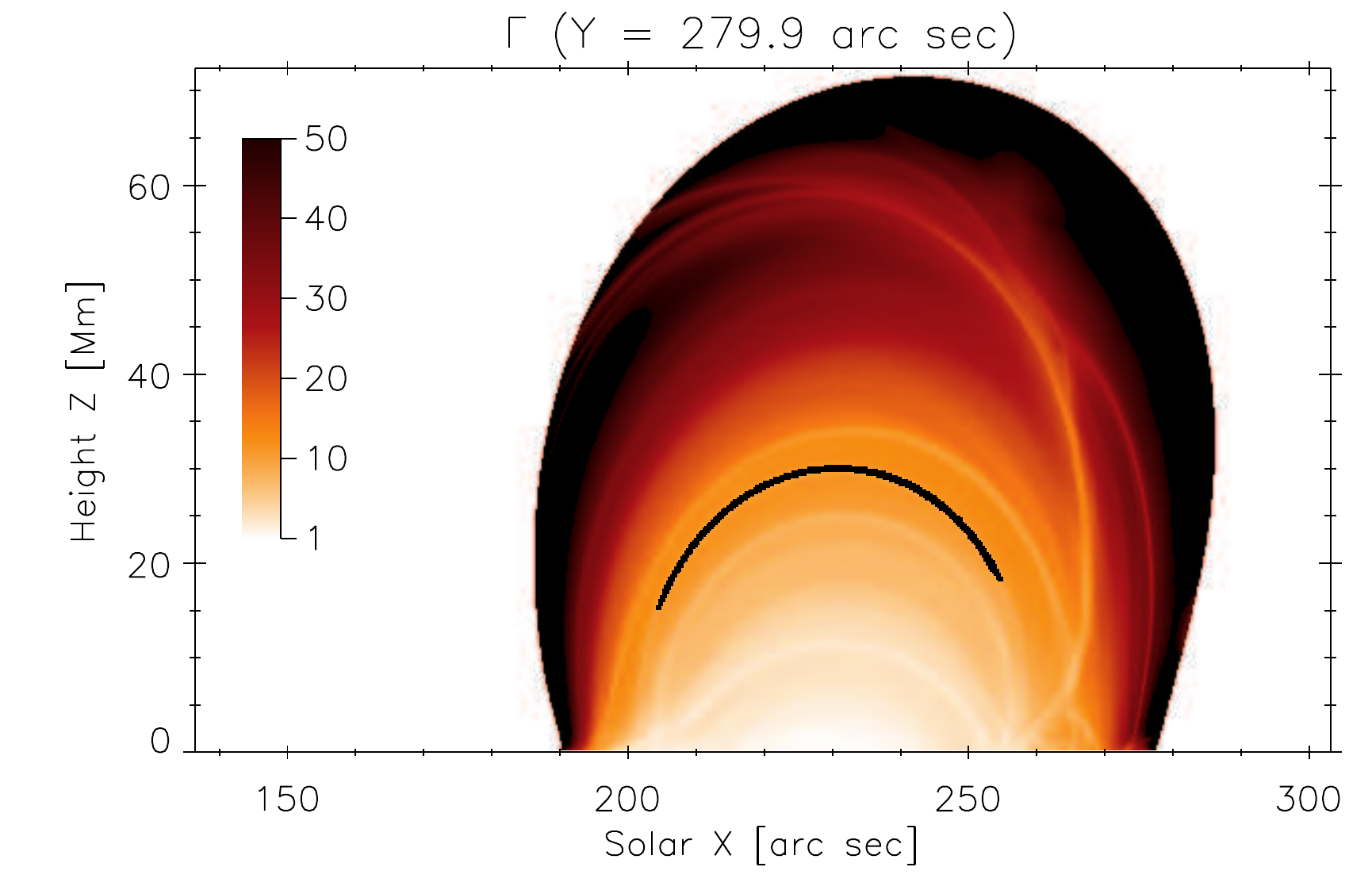}
	\includegraphics[width=8.8cm]{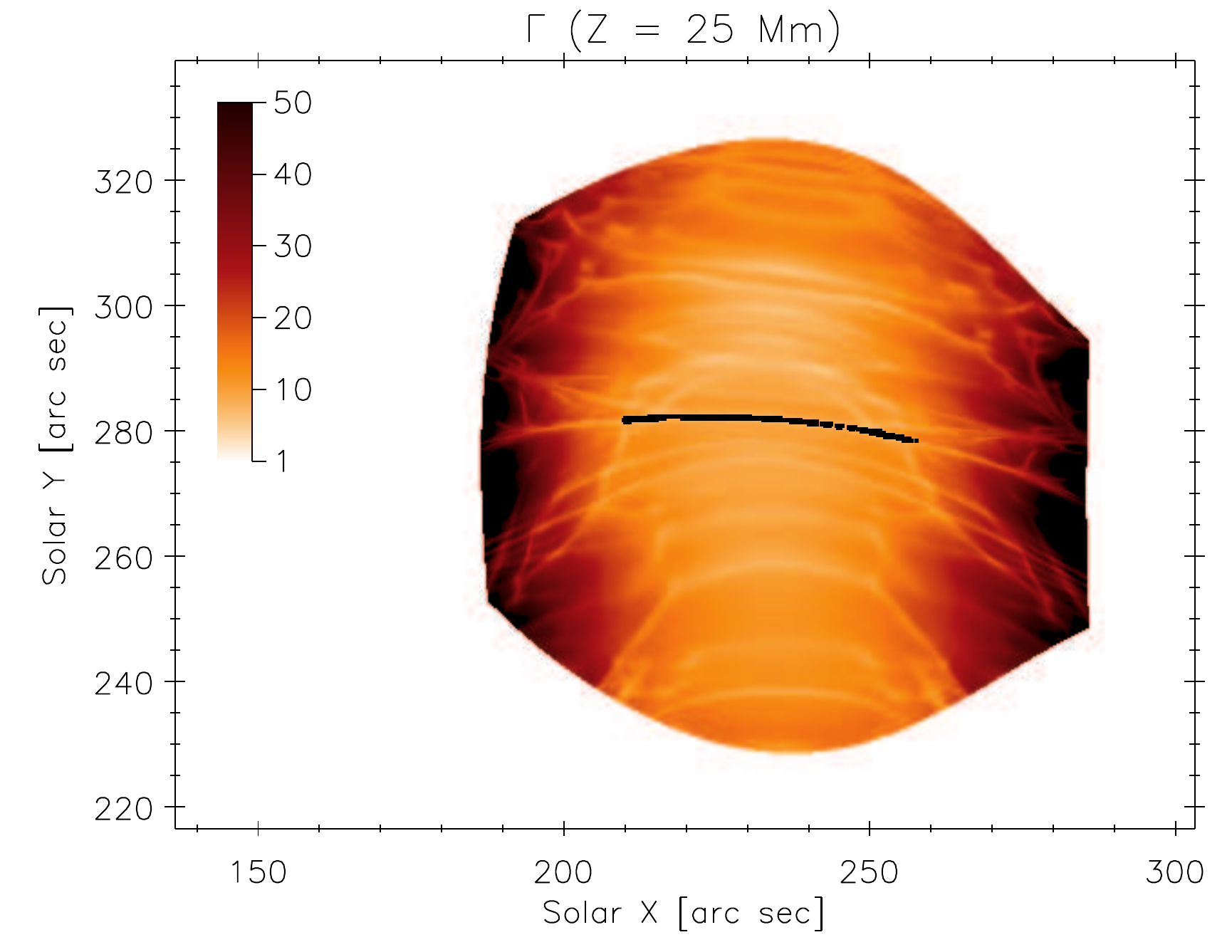}
	\includegraphics[width=16.0cm]{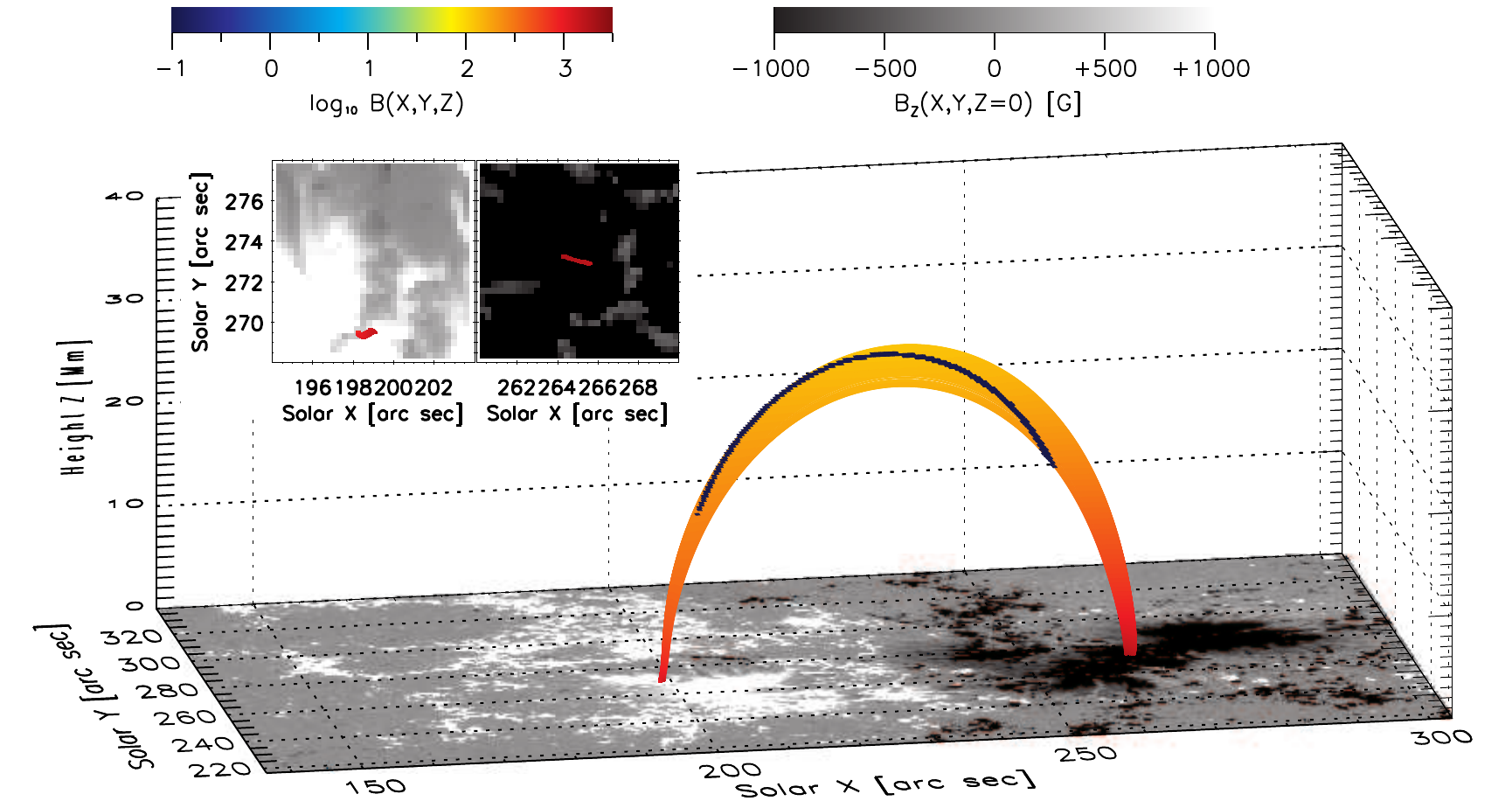}
	\includegraphics[width=16.0cm,bb=0 0 498 226,clip]{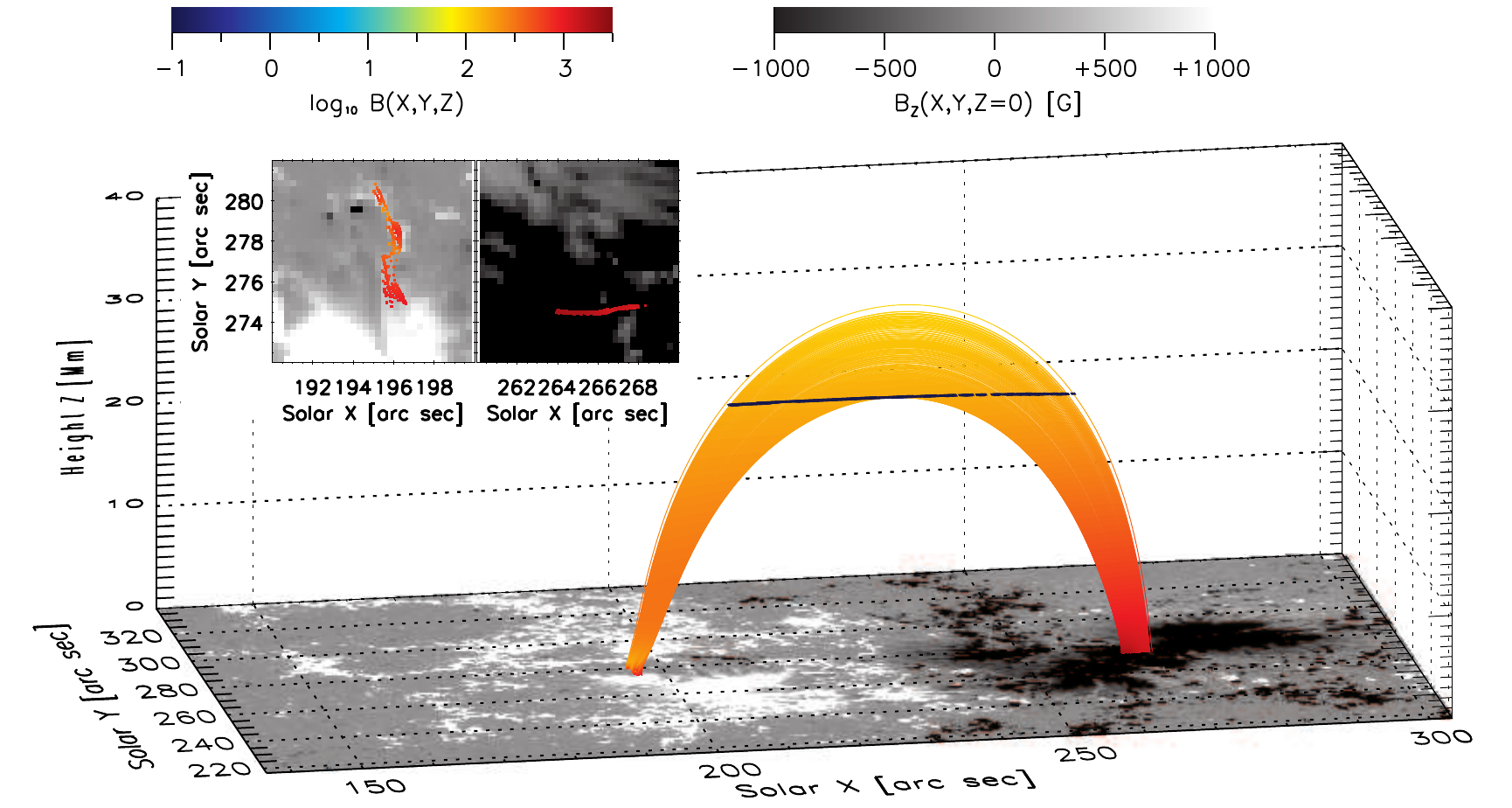}
	\caption{Selected points along two structures in cuts through $Y$\,=279.9$\arcsec$ (\textit{top left}) and $Z$\,=\,25\,Mm (\textit{top right}), as well as the 3D view of the corresponding field lines passing through these starting points (\textit{middle} and \textit{bottom}). Color denotes local values of $B$ along the field lines. The selected points are denoted as small dark squares on all images. \textit{Hinode}/SOT magnetogram is projected at the bottom boundary of the computational box. \textit{Insets}: Photospheric footpoints of the corresponding field lines.}
	\label{Fig:Gamma_FL}
   \end{figure*}
%

%
\section{Results}
\label{Sect:4}

Using the procedure outlined in Sect. \ref{Sect:3}, we calculate the area expansion factor for both the magnetic field extrapolations using the Green's function method and the submerged charges approximation. The corresponding area expansion factors are denoted $\Gamma$ and $\Gamma_C$, respectively. 

\subsection{General Characteristics of Area Expansion}
\label{Sect:4.1}

We first investigated the general characteristics of the area expansion factors. These are presented in Fig. \ref{Fig:Gamma}. The top row of this figure shows the histograms of the $\Gamma$ and $\Gamma_C$. The highest values of $\Gamma$ found are approximately 80. The histogram has a maximum at $\Gamma$\,=\,4 and then slowly decreases towards $\Gamma$\,=\,70, after which a more steep decrease is found. The $\Gamma_C$ histogram is shifted towards lower values, and exhibits a strong decrease at $\Gamma_C$\,$>$\,30. In both cases the $\Gamma_{-}$ and $\Gamma_{C,-}$ exhibits larger values than $\Gamma_{+}$ or $\Gamma_{C,+}$, respectively. This is not suprising, given that the negative magnetic flux is more concentrated in the larger sunspots than the positive-polarity flux, which originates in smaller spots and extended plage regions (Fig. \ref{Fig:SOT}, \textit{top}).

We next calculated the averaged quantities $\left<\Gamma(Z)\right>_{X,Y}$ and $\left<\Gamma_C(Z)\right>_{X,Y}$ as a function of the height $Z$. This averaging is done at each $Z$ over every pixel $[X,Y,Z$\,=\,const.$]$. Note that the number of pixels at each $Z$ is not constant, since $\Gamma$ and $\Gamma_C$ are calculated only for ``closed'' field lines (Sect. \ref{Sect:3.3}). For $Z$\,$\geq$=\,72.0\,Mm and $Z_C$\,$\geq$\,68.7\,Mm, there are no pixels containing ``closed'' magnetic field lines.

The distribution of $\left<\Gamma\right>_{X,Y}$ steadily increases from a value of $\approx$6 near the photosphere, reaching a value of $\approx$61 at $Z$\,=\,68\,Mm. The $\left<\Gamma_{-}\right>_{X,Y}$ is again larger than $\left<\Gamma_{+}\right>_{X,Y}$. Note that, unlike for the histograms, the $\left<\Gamma\right>_{X,Y}$ is by definition an exact average $\left<\Gamma_{+}\right>_{X,Y}$ and $\left<\Gamma_{-}\right>_{X,Y}$. The distribution of $\left<\Gamma_C\right>_{X,Y}$ rises more slowly to a value of $\approx$38 at $Z_C$\,=\,66.4\,Mm. The relation of $\Gamma_C$ to $\Gamma$ is discussed in Sect. \ref{Sect:4.2}.

We note that the $\left<\Gamma\right>_{X,Y}$ values are in fact lower limits. This is because of the limited size of the computational box, and thus limited number of grid points $\mathbf{r}$ containing ``closed'' magnetic field lines. This prevents the fan loops with high expansion factors, and some field lines anchored in photospheric quasi-separatrix traces \citep{Priest95,Demoulin97} to be ``closed'' in the computational box. Similarly, the decrease of $\left<\Gamma\right>_{X,Y}$ at $Z$\,$\gtrapprox$\,66 Mm is an artifact of the size of the box. 

\subsection{Spatial Distribution of Area Expansion}
\label{Sect:4.2}

We next investigated the spatial distribution of area expansion. Since the $\Gamma(\mathbf{r})$ and $\Gamma_C(\mathbf{r})$ are calculated as 3D quantities, for simplicity we show various cuts through the computational box, with one or more of the $X$, $Y$, and $Z$ coordinates fixed (Figs. \ref{Fig:Gamma_cutZ}, \ref{Fig:Gamma_cutXZ}, \ref{Fig:Gamma_cutY}, and \ref{Fig:Gamma_cutX}). Figure \ref{Fig:Gamma_cutZ} shows horizontal cuts through the computational box at three heights, $Z$\,=\,2.6, 25, and 50\,Mm. In this figure, the \textit{left} column shows the spatial distribution of $\Gamma(X,Y,Z$\,=\,const.), while the \textit{right} column shows the $\Gamma_C(X,Y,Z$\,=\,const.). We see that the $\Gamma_C$ shows a smooth spatial distribution with only a few features. However, the $\Gamma$ shows a very fine-structuring down to spatial scale of 1\,pixel corresponding to 0.3$\arcsec$. The spatial structuring in these horizontal cuts are that of many thin, thread-like structures, characterized by locally lower $\Gamma(\mathbf{r})$, embedded in smoother background of higher $\Gamma$. This structuring exists despite the fact that magnetic field becomes smooth with increasing height $Z$. For comparison, the $B_Z$ component at the height of 25\,Mm is plotted in Fig. \ref{Fig:BZ_cutZ}. The structuring in $\Gamma$ at this height (Fig. \ref{Fig:Gamma_cutZ}, \textit{left, middle}) is then caused by $\Gamma$ being a global property of each field line passing through this $Z$. 

To illuminate the spatial variation in $\Gamma$ further, in Fig.\,\ref{Fig:Gamma_cutXZ} we plot the profiles of $\Gamma$ in a series of cuts of through the active region center at $X$\,=\,230$\arcsec$ and three different heights $Z$\,=\,2.6, 25, and 50\,Mm corresponding to Fig. \ref{Fig:Gamma_cutZ}. I.e., the profiles plotted are $\Gamma(Y)$\,=\,$\Gamma(X$\,=\,230$\arcsec$,$Y, Z$\,=\,const). Note the steep local minima in $\Gamma(Y)$. The width of these minima is typically 1 to several pixels (1\,px\,$\equiv$\,0.3$\arcsec$). These local minima are separated by broad local maxima constituted by up to tens of pixels. The variation in $\Gamma_C(Y)$ is much smoother, with the steep local minima non-present. Overall, the $\Gamma_C(Y)$ is an approximate lower limit to the $\Gamma(Y)$ (Fig. \ref{Fig:Gamma_cutXZ}). Comparing the $\Gamma$ to $\Gamma_C$ we conclude that the fine-structuring of $\Gamma$ is caused by the fine-structuring of the magnetic field in the photosphere, while the overall increase of $\Gamma$ and $\Gamma_C$ with height is a result of the distribution of magnetic flux in the photosphere. 

The vertical cuts at fixed $Y$ (Fig. \ref{Fig:Gamma_cutY}) or $X$ (Fig. \ref{Fig:Gamma_cutX}) show structures reminiscent of active region loops viewed off-limb. The $Y$-cuts are performed through the centre of the active region at three different $Y$ approximately 5$\arcsec$ apart, $Y$\,=\,275.1$\arcsec$, 279.9$\arcsec$ and 285.0$\arcsec$. The increase of $\Gamma$ with height is particularly apparent in both Figs. \ref{Fig:Gamma_cutY} and \ref{Fig:Gamma_cutX}. A ``core'' at lower altitudes is surrounded by loop-like structures of width of 1 or several pixels only (Fig. \ref{Fig:Gamma_cutY}). However, these loop-like structures do not neccessarily correspond to isolated coronal loops lying along a single field-line. This is particularly apparent when considering the ``open'' loop-like structures reaching altitudes of $Z$\,$\geq$\,70\,Mm (Fig. \ref{Fig:Gamma_cutY}, \textit{top left}). We remind the reader that the $\Gamma(\mathbf{r})$ is calculated for each pixel $\mathbf{r}$ containing a field line closed within the computational box (Sect. \ref{Sect:3.3}). The apparently ``open'' loop-like structures in $\Gamma(Y$\,=\,const.) are in fact cuts through an inclined sheet of loops having similar values of $\Gamma$. 

To investigate what a particular loop-like structure in these $Y$-cuts corresponds to, we plotted field lines starting at several selected points along a single structure (Fig. \ref{Fig:Gamma_FL}). We chose starting points along a loop-like structure in the $Y$\,=\,279.9$\arcsec$ (Fig. \ref{Fig:Gamma_FL}, \textit{top left}) and along a prominent, bent thread-like structure located at approximately $Y$\,$\approx$\,280$\arcsec$ in the $Z$\,=\,25\,Mm cut (Fig. \ref{Fig:Gamma_cutZ}, \textit{middle left}, Fig. \ref{Fig:Gamma_FL}, \textit{top right}). Since manual selection of starting points would be impractical due to their large number, the selection of the starting points is done automatically based on a selected range of $\Gamma$ values (6--9 for the $Y$-cut, and 12--15 in the $Z$-cut), as well as simple geometrical constraints to avoid addition of points lying along different, nearby structures. Since the $\Gamma$ does vary both across and along these structures, these simple criteria do not lead to starting points being distributed all along the apparent structure. 

Plotting the field lines starting at the selected starting points (Fig. \ref{Fig:Gamma_FL}, \textit{middle} and \textit{bottom}) reveals that the apparent structures seen in the cuts belong to single loop-like flux-tubes. Moreover, the cross-section of these fluxtubes themselves are clearly seen to be expanding with height. The flux-tube corresponding to the loop-like structure in the $Y$\,=\,279.9$\arcsec$ cut is slightly inclined, with both footpoints being located at $Y$\,$\lessapprox$\,275$\arcsec$ in the photosphere (Fig. \ref{Fig:Gamma_FL}, \textit{middle}). The thread-like structure of $\Gamma$ in the $Z$\,=\,25\,Mm cut arises simply by cutting the corresponding loop-like flux-tube at the given height (Fig. \ref{Fig:Gamma_FL}, \textit{bottom}). 

Note that the photospheric cross-sections of these loop-like flux-tubes (insets in Fig. \ref{Fig:Gamma_FL}) are not circular, rather highly squashed. To use a food analogy, these flux-tubes resemble linguine rather than spaghetti. It is known that variations in the oblateness of the flux-tube cross-section do occur naturally from footpoint to apex \citep{Malanushenko13}. However, our results here point out that the flux-tube producing a localized structure in $\Gamma$ in fact does not have circular cross-section anywhere, neither in the photosphere nor in any cut at $Z$\,=\,const. Furthermore, the footpoints of the squashed flux-tubes are not parallel, rather, the flux-tubes exhibit a small apparent twist, about $\pi$/2 turns for the fluxtube shown in Fig. \ref{Fig:Gamma_FL}. Note that the field is by definition potential, so that no twist due to electric current is present. Rather, the apparent twist is a result of magnetic connectivity. 

%
\section{Discussion}
\label{Sect:5}

It is tempting to identify the structuring in $\Gamma$ with the structures observed in the actual solar corona. In particular, the vertical cuts are reminiscent of loops ($Y$\,=\,const.) or active regions ($X$\,=\,const.) observed off-limb, see e.g., Fig. 1 in \citet{ODwyer11}, or Fig. 18 in \citet{Malanushenko13}. In the following, we speculate on some of the implications of such an identification.

First, it is not straightforward to identify the loop-like structure in the $Y$-cut with a non-expanding observed coronal loop. As we showed in Fig. \ref{Fig:Gamma_FL}, \textit{bottom}, this is because the structure in the cut is constituted by a loop-like flux-tube whose cross-section expands with height. Nevertheless, since the (static) heating function is usually thought to be dependent both on the magnetic field and the length of the strand \citep[e.g.,][]{Mandrini00,Schrijver04,Warren06,Warren07,Lundquist08,Mok08,Dudik11}, small differences of these parameters together with small differences in $\Gamma$ (note that $\Delta\Gamma$\,=\,3 for the selected structure, Sect. \ref{Sect:4.2}) between individual strands comprising the loop-like flux-tube could lead to portioning of it into observed threads with different temperatures and densities. These threads may themselves seem to be non-expanding \citep[see Fig. 8 in][]{Peter12}. In models of coronal emission, the resulting non-expanding loops are common \citep{Mok08,Dudik11,Peter12,Lionello13}. We therefore argue that the structuring in $\Gamma$ could be responsible for the structuring of the observed solar corona. We also note that in reality, the situation will be complicated by dynamics of the heating and cooling \citep{Winebarger03,MuluMoore11,Viall11,Viall12}, so that not all strands may be visible at a given time in a given passband.

In this respect, it is important to realize that even if two strands in different flux-tubes (with different $\Gamma$) have the same magnetic fields and lengths, the difference in $\Gamma$ alone is enough to substantially change the thermodynamics of these strands \citep[see also][]{Mikic13}. This is because the volume of the two different strands, and thus the heating \textit{per particle} will be different. Therefore, the structuring of $\Gamma(\mathbf{r})$ in the coronal volume of an active region alone may significantly contribute to the observed structuring of the corona. In the Appendix, we perform simple analytical estimates of the changes in heating and electron densities for the values of $\Gamma$ derived in this paper.

Furthermore, the height distribution of $\Gamma$ could contribute to explaining the division of the active region into a hot core and a warm periphery. The active region cores, with temperatures of $log(T/$K)\,$\approx$\,6.5--6.6, have typical widths given by the separation of main photospheric polarities within the active region. Based on \textit{SDO}/AIA image in the 335\AA~passband (Fig. \ref{Fig:SOT}, \textit{bottom right}), the horizontal extent of the core along $Y$\,=\,280$\arcsec$ is about 40--50$\arcsec$, which is not untypical of active regions \citep[e.g.,][]{Warren12,UgarteUrra14,Petralia14}. It is difficult to estimate the height extent from this on-disk image. However, the height extent of the high-density (log$(n_\mathrm{e}/$cm$^{-3})$\,$\geq$\,9.5) active region core seen in \ion{Fe}{16} by e.g. \citet{ODwyer11} (Figs. 1, 6, and 9 therein) is about $\approx$30$\arcsec$, i.e., about 22\,Mm. Similar size can be obtained from e.g., Figs. 2 and 4 of \citet{Mason99}. Such horizontal and vertical sizes correspond roughly to the extent of the volume enclosed by $\Gamma$\,$\lessapprox$\,5--7. This is in line with the results of \citet{Dudik11}, who found that the modeled hot, X-ray core loops correspond to region where the magnetic field does not decrease strongly along the field lines.

These authors reported that the horizontal extent of the modeled warm EUV emission corresponds to $\Gamma$\,$\lessapprox$\,50. We also find that there are some loop-like structures in $\Gamma$ with $Z$\,$\gtrsim$\,46\,Mm (pressure scale-height for 1\,MK coronal plasma), which could correspond to the  $\approx$1\,MK loops overlying the AR core in AIA 171\AA~(Fig. \ref{Fig:SOT}, \textit{bottom left}). The loop-like structures in $\Gamma$ at heights overlying the core are typically steep local depressions in the volumetric distribution of $\Gamma$ (see Fig. \ref{Fig:Gamma_cutXZ}). These depressions are 1 to several pixels wide (Sect. \ref{Sect:4.2}). If we identify these steep local minima in $\Gamma$ as loci of coronal loops distinct from the background, the width of these minima would have important implications for the physical width of coronal loops. If the minimum is only 1 pixel wide (0.3$\arcsec$), the loop would be unresolved by the \textit{SDO}/AIA, but could be resolved by the Hi-C instrument \citep{Cirtain13,Kobayashi14}. The local minima with widths of several pixels could be resolved even by \textit{SDO}/AIA. Indeed, the observations show that there are both such examples \citep{Peter13,Brooks13,Winebarger14} and that at least some loops seen by AIA or \textit{TRACE} seem to be resolved \citep{Aschwanden05,Aschwanden11,Brooks12}. Finally, the local maxima in $\Gamma$, separating these minima, are several to several tens of arc seconds wide (Figs. \ref{Fig:Gamma_cutZ}, \ref{Fig:Gamma_cutXZ}, \ref{Fig:Gamma_cutY}, and \ref{Fig:Gamma_cutX}). We speculate that these maxima could correspond to the diffuse coronal ``background'', in which the loops are embedded \citep{Cirtain05}. Since the background is characterized by locally higher $\Gamma$, these areas could be more readily susceptible to the thermal nonequilibrium \citep{Mikic13} producing the ubiquitious coronal rain \citep{Antolin12}, although we cannot exclude the presence of this phenomenon also in the loops \citep{Dudik11}, if their $\Gamma$ is high enough.

Note that in our calculation, the magnetic field is assumed to be potential and thus free of the Lorentz force. Therefore, the structures seen in $\Gamma(\mathbf{r})$ are in force balance and must be native to even the potential extrapolated magnetic fields. It is not clear whether these structures could also have counterparts in the non-linear force-free fields. However, it is known that coronal currents can produce additional complexity in magnetic connectivity \citep[e.g.,][]{Buchner06,Aulanier12,Savcheva12}, and therefore it is likely that the non-linear force-free fields will also exhibit structuring in $\Gamma$. However, this point should be investigated in the future. We remind the reader that due to the potential approximation used here, the results presented in this paper are valid only for loops with no twist and active regions that do not produce significant flares.

We caution that both the potential and non-linear force-free fields obtained by matching the magnetic field to the stereoscopically inferred geometry of the coronal loops \citep[e.g.,][]{Aschwanden10,Sandman11,Aschwanden13b,Aschwanden13a,Aschwanden14,Gary14} rely typically on 10$^2$ or less submerged magnetic charges or dipoles. Since the approximation by submerged charges retain only the most pronounced structures in the expansion factor (see Figs. \ref{Fig:Gamma_cutZ} and \ref{Fig:Gamma_cutX}), the structuring of the expansion factor in stereoscopically constrained magnetic field models relying on approximations by submerged charges or dipoles may possibly be underestimated and/or lower than found here for potential fields. This question is however out of the scope of this paper. More work is needed to investigate this problem.

The absence of the fine-scale structuring in the $\Gamma_C$ found in Sect. \ref{Sect:4.2} can be further illuminated if the approximation by submerged charges is thought of as an analogue to retaining only several of the lowest Fourier harmonics in the Fourier-transformed magnetogram. Adding progressively higher Fourier harmonics (until the magnetogram is reconstructed) will lead to the fine-scale structure in both the magnetogram, the extrapolated field \citep{Alissandrakis81,Gary89}, as well as the volumetric distribution of the expansion factor $\Gamma$ \citep{Dudik11}. An important caveat however is that the Fourier-transform method can be used only to extrapolate flux-balanced magnetograms \citep{Alissandrakis81,Gary89}. If the magnetogram is not flux-balances, as is the case here (Sects. \ref{Sect:2} and \ref{Sect:3.1}), the Fourier-transform method would enforce the flux-balance by setting the lowest Fourier harmonic to zero. Since our direct extrapolation (Eq. 1) circumvents the flux-balance assumption, the $\Gamma$ calculated by the Eq. (1) and Fourier transform method will not be exactly the same.

Since we are using a potential extrapolation, we note that kinetic pressure gradients across the field, graviational stratification, and plasma flows are not considered. However, these forces are unlikely to contribute to the observed non-expansion of coronal loops in a low-$\beta$ medium \citep{Petrie06,Petrie08}, but may affect the behaviour of the field expansion in the non-force-free regions above the photosphere and in chromosphere \citep{Metcalf95}. In principle, separating these effects form the expansion of the field in a low-$\beta$ corona can be done if measurements of the magnetic fields in the upper chromosphere are used instead of a photospheric magnetogram. However, based on our results, we expect that the fine-structuring of the expansion factor in potential fields calculated by extrapolating a chromospheric magnetogram would still correspond to the fine-structuring in such a magnetogram.

Finally we note that our definition of $\Gamma$ (Eq. \ref{Eq:Gamma}) differs from those of \citet{Asgari12} and \citet{Asgari13}. These authors define the expansion factor as $\Gamma_\mathrm{cor}$\,=\,$B_\mathrm{TR}/B_\mathrm{min}$, i.e., a ratio of the magnetic field in the transition region to the minimum of the magnetic field along the given field line. Note that (1) the $B_\mathrm{TR}$ can be model-dependent, and (2) the the minimum of the magnetic field can in principle occur at different spatial location than $B_Z$\,=\,0 (apex, used in Eq. \ref{Eq:Gamma}). However, we point out that the values of $\Gamma_\mathrm{cor}$ obtained by \citet{Asgari13} (Table 1 therein) are similar to values obtained here. Furthermore, judging from Fig. \ref{Fig:Gamma_FL}, the minimum of the magnetic field is located reasonably close to the apices of the plotted field lines.

%
\section{Summary}
\label{Sect:6}

We performed a Green's function extrapolation of a photospheric longitudinal magnetogram of a quiescent active region NOAA 11482 observed by \textit{Hinode}/SOT-SP. We also performed an approximation of the magnetogram with 134 submerged charges. This approximation allows the retention of the observed flux distribution, while destroying the small-scale structure. Both potential magnetic fields are similar.

From these magnetic fields, we calculated the area expansion factor of a strand passing through a given 3D grid point. We found that on average, the expansion factor rises with height to values over 50. The spatial distribution exhibits significant structure on spatial scales of 1 or several pixels, where 1\,pixel\,=\,0.3$\arcsec$. This fine-scale structure is missing in the magnetic field calculated from the distribution of submerged charges. The magnetic field of the submerged charges show smooth spatial variations of its area expansion factors, which are an approximate lower limits to the expansion factors of the magnetic field obtained by our extrapolation of the SOT magnetogram.

In vertical cuts, the spatial distribution of the expansion factor shows loop-like structures with locally lower $\Gamma$. These structures are created by loop-like flux-tubes with expanding, highly squashed cross-sections. These loop-like structures are embedded in a background corresponding to smoother variations and higher values of the expansion factor. We argued that such structuring of the expansion factor could be responsible for the creation of the observed coronal loops by virtue of changing the heating per particle. 

In summary, we showed that the potential magnetic fields calculated by direct extrapolation of observed magnetic fields in the solar photosphere possess a fine-structure in the area expansion factor. Such potential fields are composed of many flux-tubes distinct from their neighborhood, resembling linguine rather than spaghetti.

\acknowledgements
The authors thank G. Del Zanna and H. E. Mason for useful discussions. AIA and HMI data are courtesy of NASA/SDO and the respective science teams. Hinode is a Japanese mission developed and launched by ISAS/JAXA, with NAOJ as domestic partner and NASA and STFC (UK) as international partners. It is operated by these agencies in co-operation with ESA and NSC (Norway). JD acknowledges support from the Newton Fellowship Programme (Royal Society). EDZ acknowledges the support by Grant Agency of the Czech Republic, Grant No. P209/12/1652.

\bibliographystyle{apj}
\bibliography{Extrapol}

\appendix
\section{Heating and Density in an Expanding Loop Strand}
\label{Appendix:A}

\subsection{Toy Model for an Expanding Coronal Loop Strand}
\label{Appendix:A1}

In this appendix, we estimate the changes in total heating and resulting electron density in an expanding loop. To do that, as well as for purposes of analytical tractability, we build a simple toy model based on several simplifying assumptions. First, we assume a non-inclined, semi-circular loop strand with an expanding cross-section, similar to that of an individual field-line shown in Fig. \ref{Fig:Gamma_FL}. We denote $A_0$ the (infinitesimal) photospheric cross-section of this strand, $s$ the coordinate along the strand, and $L$ the half-length of the strand, i.e., the total footpoint-to-apex distance. Furthermore, we assume that the magnetic field $B(s)$ along this strand is decreasing exponentially with a scale-lenght $s_B$, i.e., 
\begin{equation}
  B(s) = B_0 *\mathrm{e}^{-s/s_B}\,,
  \label{Eq:B(s)}
\end{equation}
and that the volumetric heating rate $E_\mathrm{H}(s)$ is steady, non-uniform and depends \textit{locally} on some power $\alpha > 0$ of the magnetic field, i.e.,
\begin{equation}
  E_\mathrm{H}(s) = B_0^\alpha *\mathrm{e}^{-\alpha s/s_B}\,,
  \label{Eq:E_H(s)}
\end{equation}
where we the quantity the $s_\mathrm{H} = s_B/\alpha$ is the heating scale-length. The assumption (\ref{Eq:B(s)}) is justified in potential fields, where the field decreases strongly along a field line, especially for field lines rooted in strong photospheric flux concentrations \citep{Dudik11}. The assumption of the heating depending on the magnetic field is a common one in modeling the active region emission \citep[e.g.,][]{Mandrini00,Schrijver04,Warren06,Warren07,Lundquist08,Mok08,Dudik11,Mikic13}.

The magnetic field profile (\ref{Eq:B(s)}) will produce an exponentially increasing cross-section $A(s)$. with a total expansion factor of the strand $\Gamma$ given by 
\begin{equation}
  \Gamma = B_0 /B(s=L) = \mathrm{e}^{L/s_B}\,.
  \label{Eq:Gamma_total}
\end{equation}
The volume $V(\Gamma)$ of the strand can then be obtained by integrating $A(s)$ along the strand. Since the strand is semi-circular, the location $s$ can be substituted by the expression $s = 2L\phi/\pi$, where $\phi$\,$\in$\,$\left<0,\pi/2\right>$ is an angular variable along the strand. The cross-section $A(\phi)$ is then $A(\phi) = A_0 *\mathrm{e}^{2\phi \mathrm{ln}(\Gamma)/\pi}$. We get
\begin{equation}
  V(\Gamma) = \int_{0}^{\pi/2} A(\phi) \frac{2L}{\pi} d\phi = A_0 L\frac{\Gamma-1}{\mathrm{ln}(\Gamma)}\,.
  \label{Eq:V_total}
\end{equation}
Similarly, the total heating input $H$ in this volume is obtained as
\begin{equation}
  H(\Gamma) = \int_{0}^{\pi/2} B_0^\alpha \mathrm{e}^{-2\alpha L\phi/\pi s_B} A(\phi) \frac{2L}{\pi} d\phi =  A_0 L B_0^\alpha \frac{1-\Gamma^{1-\alpha}}{(\alpha-1)\mathrm{ln}(\Gamma)}\,,
  \label{Eq:H_total}
\end{equation}
where the identity (\ref{Eq:Gamma_total}) have been used. The expression (\ref{Eq:H_total}) is valid for $\alpha$\,$\ne$\,1; for $\alpha$\,=\,1 we obtain $H(\Gamma)$\,=\,$A_0 L B_0$, since $E_\mathrm{H}(s) A(s)$\,=\,$A_0 B_0$ for $\alpha$\,=\,1.

   \begin{figure}[!ht]
	\centering
	\includegraphics[width=8.8cm]{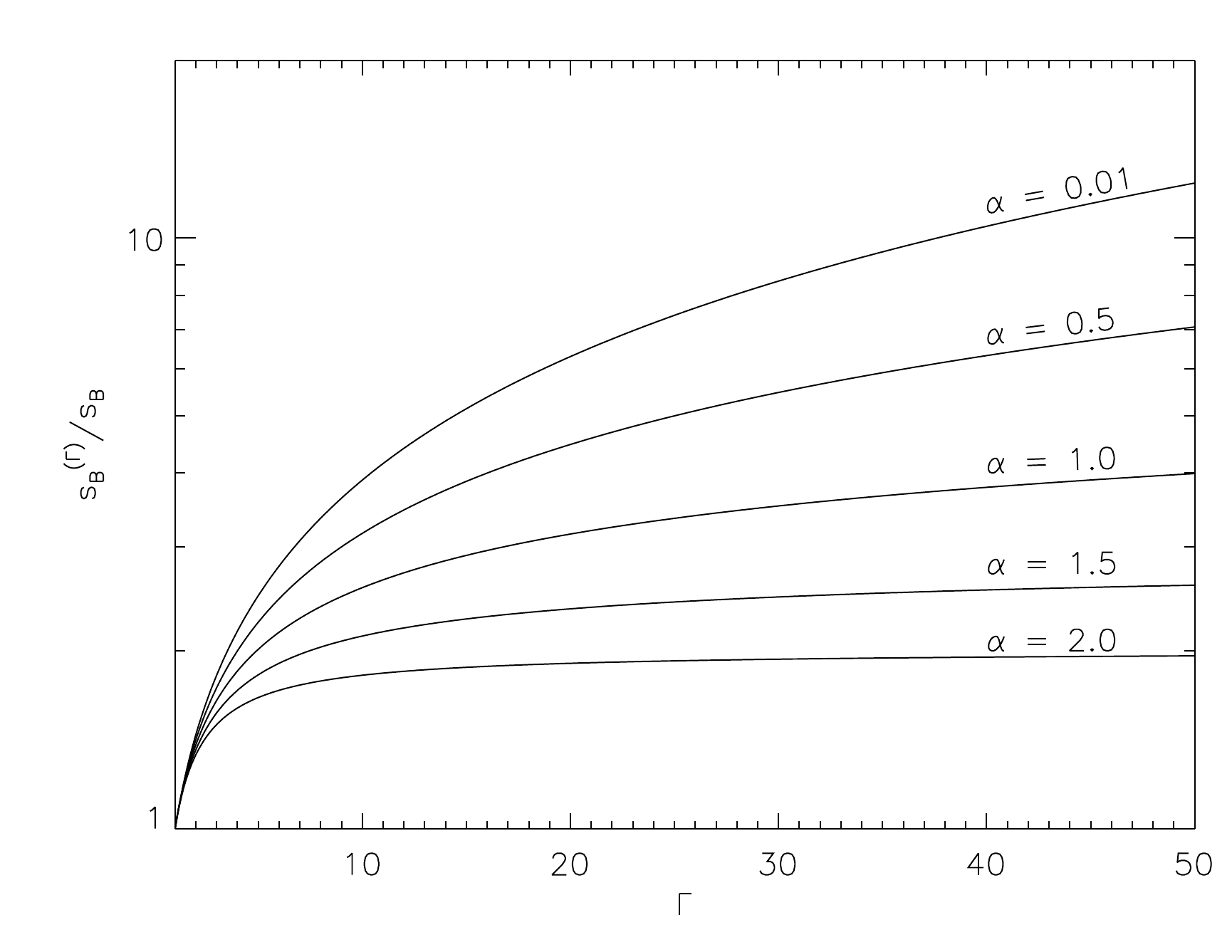}
	\caption{Equivalent heating scale-length for a non-expanding strand having the same total heat input as the strand with an expanding cross-section (see text for details).}
	\label{Fig:A1}
   \end{figure}
%

\subsection{Equivalent Heating in a Non-Expanding Strand}
\label{Appendix:A2}

Note that the formula (\ref{Eq:H_total}) is derived in a self-consistent manner: The volumetric heating rate, as well as the cross-section, both depend directly on the magnetic field given by expression (\ref{Eq:B(s)}). This is an important point, since the assumption of a constant strand cross-section even when the magnetic field is obtained from an extrapolation of a photospheric magnetogram \citep[e.g.,][]{Warren06,Warren07} may be misleading. We illustrate this point by calculating the total heating input $H^{*}$ into a non-expanding strand ($A(s) = A_0$) with the same non-uniform heating
\begin{equation}
  H^{*}_{A(s) = A_0} =  A_0 B_0^\alpha \frac{s_B}{\alpha} \left(1-\frac{1}{\Gamma^{\alpha}}\right)\,,
  \label{Eq:H1_total}
\end{equation}
where the identity (\ref{Eq:Gamma_total}) have again been used. Note that $H^{*}(s_B,\Gamma) < H(\Gamma)$. To get a measure of how much the $H^{*}$ underestimates the $H(\Gamma)$, we define an ``equivalent'' $s_B^{(\Gamma)}$, for which $H^{*}\left(s_B^{(\Gamma)},\Gamma\right) = H(\Gamma)$, leading to 
\begin{equation}
  s_B^{(\Gamma)} = s_B \frac{\alpha}{\alpha-1} \frac{1-\Gamma^{1-\alpha}}{1-\Gamma^{-\alpha}}\,,
  \label{Eq:sB_equiv} 
\end{equation}
valid for $\alpha$\,$\ne$\,1. For $\alpha$\,=\,1 the following expression holds 
\begin{equation}
  s_B^{(\Gamma)} = s_B \frac{\mathrm{ln}(\Gamma)}{1-1/\Gamma}\,.
  \label{Eq:sB_equiv_alpha1} 
\end{equation}
The ratio $s_B^{(\Gamma)}/s_B$ is plotted in Fig. \ref{Fig:A1}. We see that $H^{*}$ given by Eq. (\ref{Eq:H1_total}) significantly underestimates the $H(\Gamma)$ even for small values of $\Gamma$\,$\to$\,1. Note that this is because the strongest expansion occurs close to photosphere (see Eq. \ref{Eq:B(s)}), where the heating is also assumed to be the strongest.

\subsection{Density Structure of an Atmosphere with Structured Area Expansion}
\label{Appendix:A3}

Having found a significant structure in the distribution of area expansion $\Gamma$ in Sect. \ref{Sect:4.2}, we next estimate the increase of electron density $n_\mathrm{e}$ in a flux-tube that is characterized by lower $\Gamma$ than the neighbouring regions (see Figs. \ref{Fig:Gamma_cutZ}, \ref{Fig:Gamma_cutXZ}, \ref{Fig:Gamma_cutY}, and \ref{Fig:Gamma_cutX}). To do that, we assume that the strands are in equilibrium with a steady, non-uniform heating given by Eq. (\ref{Eq:E_H(s)}). Under such conditions, the strand atmosphere can be described by scaling laws \citep[e.g.,][]{Serio81,Aschwanden02,Dudik09,Dudik11}. Neglecting the pressure variations with height and assuming the power-law radiative-loss function with a slope of $-$1/2, the electron density scales with loop parameters as
\begin{equation}
  n_\mathrm{e} = C B_0^{4/7} L^{1/7} \mathrm{e}^{-3\gamma_1 L/s_\mathrm{H}} \mathrm{e}^{-2\gamma_2 L/7s_\mathrm{H}}\,,
  \label{Eq:ne}
\end{equation}
\citep[c.f., Eqs. (25) and (26) in][]{Dudik09}. In this expression, $\gamma_1$\,$\approx$\,$-0.09$ and $\gamma_2$\,$\approx$\,0.7 are parameters \citep[Table 1 in]{Dudik09,Dudik11}, and $C$ is a constant independent of $B_0$, $L$ and $s_\mathrm{H}$. 

Substituting the expression (\ref{Eq:Gamma_total}) together with $s_\mathrm{H} = s_B/\alpha$, we obtain
\begin{equation}
  n_\mathrm{e} = C B_0^{4\alpha/7} L^{1/7} \Gamma^{-3\alpha\gamma_1 -2\alpha\gamma_2 /7}\,,
  \label{Eq:ne_Gamma}
\end{equation}
i.e., a corona characterized by a highly structured expansion factors $\Gamma$ will also be characterized by a highly structured density \citep[see also][]{Dudik11}. This expression permits an estimate of the density increase as a function of $\Gamma$. As an illustration, suppose that we have two loop strands, characterized by the same $L$ and $B_0$, but with different expansion factors $\Gamma_1$\,=\,8 and $\Gamma_2$\,=\,21. The value of $\Gamma_1$\,=\,8 corresponds to a typical minimum values found in cut through the height $Z$\,=\,25\,Mm and $X$\,=\,230$\arcsec$ (Fig. \ref{Fig:Gamma_cutXZ}, \textit{middle}), while $\Gamma_2$\,=\,21 corresponds to the typical mean value of $\Gamma$ at this height $Z$ (Fig. \ref{Fig:Gamma}, \textit{bottom left}). The corresponding density ratios of these two strands would then be
\begin{equation}
  \frac{n_\mathrm{e,1}}{n_\mathrm{e,2}} = \left(\frac{\Gamma_1}{\Gamma_2}\right)^{-3\alpha\gamma_1 -2\alpha\gamma_2 /7} \approx (1.57)^\alpha\,,
  \label{Eq:ne_ratio}
\end{equation}
corresponding to an increase of emission measure of $(1.57)^{2\alpha}$\,$\approx$\,(2.48)$^\alpha$ of the strand 1 compared to the strand 2. The corresponding ratio of the heating \textit{per particle} can be obtained as (c.f., Eqs. \ref{Eq:V_total} and \ref{Eq:H_total})
\begin{equation}
  \frac{H_1}{n_\mathrm{e,1} V_1}\left(\frac{H_2}{n_\mathrm{e,2} V_2}\right)^{-1} = \left(\frac{\Gamma_1}{\Gamma_2}\right)^{3\gamma_1 +2\gamma_2 /7} \frac{(\Gamma_2-1) \mathrm{ln}(\Gamma_1)}{(\Gamma_1-1) \mathrm{ln}(\Gamma_2)}  \approx 1.24\,
  \label{Eq:H_ratio}
\end{equation}
for $\alpha$\,=\,1 (with $H_1$\,=\,$H_2$), and 
\begin{equation}
  \frac{H_1}{n_\mathrm{e,1} V_1}\left(\frac{H_2}{n_\mathrm{e,2} V_2}\right)^{-1} = \left(\frac{\Gamma_1}{\Gamma_2}\right)^{3\alpha\gamma_1 +2\alpha\gamma_2 /7} \frac{(\Gamma_2-1)\left(1-\Gamma_1^{1-\alpha}\right)}{(\Gamma_1-1)\left(1-\Gamma_2^{1-\alpha}\right)}\,
  \label{Eq:H_ratio_alpha}
\end{equation}
for $\alpha$\,$\ne$\,1.


\end{document}